\documentclass[twocolumn]{aastex631}

\usepackage{graphicx}	
\usepackage{float}      
\usepackage{amsmath}	
\usepackage{amsfonts}
\usepackage{cancel}
\usepackage{amsfonts}
\usepackage{placeins}

\usepackage{hyperref}
\hypersetup{
    colorlinks=true,     
    linkcolor=blue,      
    citecolor=blue,      
    urlcolor=blue        
}

\usepackage{fancyhdr}
\begin{document}
\title{OTI on FIRE: Testing the Efficacy of Orbital Torus Imaging to Recover the Galactic Potential}

\correspondingauthor{Micah Oeur}
\email{moeur@ucmerced.edu}

\author[0000-0001-5636-3108]{Micah Oeur}
\affiliation{Department of Physics, University of California, Merced \\
5200 Lake Road, Merced, CA 95343, USA}

\author[0000-0003-3217-5967]{Sarah R. Loebman}
\affiliation{Department of Physics, University of California, Merced \\
5200 Lake Road, Merced, CA 95343, USA}

\author[0000-0003-0872-7098]{Adrian M. Price-Whelan}
\affiliation{Center for Computational Astrophysics, Flatiron Institute, \\
162 Fifth Ave, New York, NY 10010, USA}

\author[0000-0002-8354-7356]{Arpit Arora}
\affiliation{Department of Astronomy, University of Washington, Seattle, WA 98195, USA}

\author[0000-0003-2806-1414]{Lina Necib}
\affiliation{Department of Physics and MIT Kavli Institute for Astrophysics and Space Research, \\
77 Massachusetts Avenue, Cambridge, MA 02139, USA}
\affiliation{The NSF AI Institute for Artificial Intelligence and Fundamental Interactions, \\
77 Massachusetts Avenue, Cambridge, MA 02139, USA}

\author[0000-0003-1856-2151]{Danny Horta}
\affiliation{Center for Computational Astrophysics, Flatiron Institute, \\
162 Fifth Ave, New York, NY 10010, USA}

\begin{abstract}

Orbital Torus Imaging (OTI) is a dynamical inference method for determining the Milky Way's gravitational potential using stellar survey data. 
OTI uses gradients in stellar astrophysical quantities, such as element abundances, as functions of dynamical quantities, like orbital actions or energy, to estimate the Galactic mass distribution, assuming axisymmetry and steady-state of the system. While preliminary applications have shown promising outcomes, its sensitivity to disequilibrium effects is unknown. 
Here, we apply OTI to a benchmark Feedback in Realistic Environments (FIRE-2) cosmological hydrodynamic simulation, m12i, which enables a comparative analysis between known FIRE-2 vertical acceleration profiles and total surface mass densities to the analogous OTI-inferred results. 
We quantify OTI’s accuracy within 16 solar-analog volumes embedded in the simulated galactic disk. 
Despite a dynamically-evolving system, we find that OTI recovers the known vertical acceleration profiles within 3$\sigma$/1$\sigma$ errors for 94\%/75\% of the volumes considered.
We discuss the method's sensitivity to the local, instantaneous structure of the disk, reporting a loss in accuracy for volumes that have large ($>$1.5 kpc) scale heights and low total density at $z$=1.1 kpc. 
We present realistic OTI error bars from both MCMC sampling and bootstrapping the FIRE-2 simulated data, which provides a touchstone for interpreting results obtained from current and forthcoming surveys such as SDSS-V, Gaia, WEAVE, and 4MOST. 

\end{abstract}

\section{Introduction} 
\label{sec:intro}

Understanding the internal dynamics of galaxies and accurately modeling their gravitational potential is essential to constrain the nature of dark matter (DM) and alternative theories of gravity \citep[e.g.,][]{Read2014, Chakrabarti2020, Banik22, Zentner22}. DM does not directly interact with light but is detectable through its gravitational effect on stellar and gas velocities \citep{RF70, Bosma81, Eilers2019, Buschmann2021, Ou24, Roche24}.  Thanks to recent surveys like Gaia \citep{Gaia16}, large stellar datasets of 5D and 6D phase-space information are now available for the Milky Way (MW). This enables detailed studies of the mass distribution, providing a unique perspective on the nature of DM compared to other galactic systems. 

Historically, MW models have relied on simplified mathematical frameworks to connect stellar velocities to the underlying DM distribution \citep{Kapteyn1922, Oort1932, Bahcall1980, Kuijken1989a}. These approaches, which assume symmetries in the gravitational potential or simplifications of the distribution function, remain widely used to constrain the local DM density distribution \citep[e.g.,][]{Zhang2013, McKee2015, Guo2020}. Many studies make use of the vertical Jeans equation \citep{Jeans1915} to connect the kinematics of stars in the solar neighborhood to the underlying DM distribution through the use of Poisson's equation. Results from these studies are generally consistent with those derived from extrapolating the MW's rotation curve \citep[e.g.,][]{DehnenBinney1998a, Huang2016, Eilers2019, Ou24, Bienayme24, Koop2024}. 
However, there are differences in the errors and overlap of the results found for the local DM density \citep[e.g.][]{Bienayme2014, Huang2016, Eilers2019}, motivating constraining the DM density with other methods. 

Why use simplified models to constrain the DM distribution? The value of these methods lies in their well-defined mathematical frameworks that offer interpretable results that are directly connected to observables. For example, traditional approaches, such as the Jeans equation and Schwarzschild modeling \citep{Jeans1922, Kuijken1989, Schwarzschild93} have been extensively employed to estimate the Galactic acceleration and infer the total density of the MW \citep{Bahcall1984,  VDM91, Watkins2010}. These models also serve as a baseline for interpreting and understanding more complicated $N$--body models. In return, these more sophisticated simulations offer a way to test inferences derived from closed-form analytic models against fully characterized systems \citep{Loebman2012, Loebman2014, Koop2024}. This allows us to determine whether departures from simplifying assumptions -- such as the MW being in steady-state, axisymmetric, and containing homogeneous stellar populations -- significantly impact dark matter estimates \citep{Ou25}. Several studies \citep{vasiliev21, dsouza22, Arora22, Arora2024c} have shown that relaxing these assumptions is critical for orbit modeling and to get an accurate mass distribution measurement of the MW halo globally.

These methods are challenged by time-dependent external perturbations \citep[like interactions with satellite galaxies, e.g. Sagittarius Dwarf Spheroidal or the LMC,][]{Abadi2003, Belokurov2006, Laporte18}, which drive the system out of equilibrium, indicating a need for models that may be robust to non-equilibrium dynamics and/or an understanding of how disequilibrium impacts dynamical inference results. Additionally, we see nonaxisymmetric structures in the disk, including the bar and spiral arms, which can generate the moving groups, and phase-space spirals observed in the MW \citep{Gomez16, Laporte18, Antoja2018, Helmi_2018, Chiba2021}, complicating the validity of steady-state assumptions. 

While time-dependent external perturbations and internal features challenge the assumption of equilibrium in the MW, simulations have been invaluable in exploring these disequilibrium effects and understanding the galactic DM distribution.  
Simulations also allow an investigation of how DM profiles are influenced by mergers, substructures, and interactions. As shown in \citet{Gomez16} and \citet{Garavito-Camargo2019}, not only is the galactic disk changed, but so is the distribution of DM in the galactic halo when a satellite infalls. The passage of the LMC can excite wakes in the DM halo, leading to correlated dynamics between the DM halo and the stellar disk. Additional ways in which satellites have impacted the dynamics in the disk have been explored by \citet{Purcell2011}, who found the Sagittarius impact could induce spiral arms and ring-like features in the outer disk, changing how mass is distributed in the MW. \citet{Widrow2014} demonstrated that a satellite can transfer some of its energy to disk stars, resulting in coherent stellar motion, such as bending and breathing modes. Without a doubt, accretion of satellite galaxies can have a profound effect on stellar motion in the disk. Additionally, \citet{Deason2013}, using the suite of simulations from \citet{Bullock2005}, studied the impact of accretion events on stellar halos and found breaks in the stellar and DM density profiles correspond to satellite galaxy interactions. \citet{Arora2024} showed, using FIRE-2 simulations, that an LMC analog boosts encounter rates between DM subhalos and stellar streams by 10-40$\%$ near the satellite, due to LMC analog subhalos, an induced density wake in the host’s DM halo, and an induced host reflex motion. Furthermore, \citet{arora2025} demonstrated that this satellite-induced response is coupled with the accretion history and true shape of the DM halo. These studies further emphasize the complex relationship between stellar and DM distributions. 

In spite of the complexities of the non-axisymmetric DM distribution, tools like Orbital Torus Imaging (OTI) \citep{APW21, Horta2024} hold promise to yield robust estimations of the underlying potential. OTI assumes that the stellar distribution function is symmetric in phase space and the vertical kinematics decouple from in-plane kinematics (which are often satisfied in steady-state and axisymmetric systems). It leverages gradients in stellar properties (e.g., age, chemical abundance) as functions of dynamical quantities (e.g., orbital actions, energy) to estimate the galactic mass distribution. Advantages of the model include its insensitivity to the selection function and that it sidesteps both the need for a parameterized potential model and to integrate orbits for a large number of stars. \cite{pricewhelan2024datadriven} applied OTI to the final snapshot of an $N$--body merger simulation between a dwarf galaxy perturber and a disk galaxy host \citep{Hunt2021}, finding only a $\sim$15$\%$ offset between predictions from OTI and the simulation's ground truth. While this analysis demonstrates OTI's prospects in yielding accurate estimates even in non-equilibrium systems, it also opens the possibility of extending OTI analysis to systems with gas dynamics and feedback mechanisms, addressing a wider range of astrophysical environments and enhancing the robustness of OTI-derived galactic mass distribution estimates.

Advances in computing technology have brought about sophisticated hydrodynamic simulations that trace the evolution of galaxies within a cosmological context \citep{Vogelsberger2014, Crain2015, Schaye2015}. High-resolution zoom-in simulations including baryonic physics, like AURIGA \citep{Grand2017} and the \textit{Latte} suite of Feedback in Realistic Environments (FIRE-2) galaxies \citep{Wetzel2016, hopkins_2018_fire2}, have provided realistic models of galaxy formation and evolution, allowing us to test new techniques to model the galactic potential \citep{Arora22, Rose2024} and significantly enhance our understanding of the structure and distribution of DM within galaxies.

Recently, \cite{Necib2019} found a correlation between the accreted stellar and DM velocity distributions from luminous satellites within the Solar circle in the FIRE-2 simulations, showcasing how mergers influence local DM features. Similarly, \cite{Arora2024} compared cold dark matter (CDM) and self-interacting dark matter (SIDM) halos in the FIRE-2 simulations, revealing density profile differences and steeper SIDM vertical acceleration gradients near the Solar Neighborhood. In this context, OTI can help constrain different dark matter models by providing detailed estimates of the galactic mass distribution, including contributions from DM. While \citet{pricewhelan2024datadriven} explored an isolated merger event, we apply OTI to the \textit{Latte} suite of MW-mass FIRE-2 simulations \citep{hopkins_2018_fire2, Wetzel2016}, which includes full hydrodynamics. These simulated galaxies form complex dynamical structures, like bars and spiral arms, and include hierarchical assembly through mergers. 

While OTI has provided estimates of potential parameters for the MW \citep{APW21, Horta2024}, how OTI inferences are impacted by disequilibrium has not been fully characterized. Our implementation of OTI on FIRE-2 simulation m12i—a recently quiescent MW-mass galaxy—aims to assess how disequilibrium impacts the recovery of potential parameters, like the vertical acceleration. 
In this paper, we utilize the stellar acceleration data and other tracked properties (like stellar positions, velocities, and chemical abundances) from the \textit{Latte} suite of FIRE-2 cosmological zoom-in simulations to quantify the impact of disequilibrium on potential parameter recovery by comparing OTI-inferred parameters to FIRE-2 simulated ground truth. 

The structure of the paper is as follows:  \S\ref{sec:simulations} describes the FIRE-2 simulated datasets, \S\ref{sec:methods} outlines the OTI methodology,  \S\ref{sec:application} details the application of OTI to FIRE-2 data, \S\ref{sec:results} presents our results, \S\ref{sec:disc} offers a discussion, and \S\ref{sec:conclusion} concludes the paper.

\section{The \textit{Latte} Simulations}\label{sec:simulations}
In this section we discuss m12i, the cosmological simulation we utilize in this paper. Initial conditions for m12i were drawn from the AGORA project \citep{Kim2014} assuming the following cosmology \citep{Wetzel2016}: $h = 0.702$, $\Omega_{\Lambda} = 0.728$, $\Omega_{m} = 0.272,$, $\Omega_{b} = 0.0455$, $\sigma_{8}= 0.807$ and $n_{s} = 0.961$.  We select the halo, m12i, as it is a recently quiescent  \citep{Panithanpaisal21, Garavito-Camargo24} MW-mass galaxy whose properties have been well-characterized \citep[][]{Wetzel2016, Necib2019, Sanderson2020, Carillo2023, Wetzel2023, Bhattarai2024}.

We simulate m12i using the Feedback in Realistic Environments (FIRE-2)  \citep{hopkins_2015_mesh_free_hydrodynamic_simulations, hopkins_2018_fire2, Wetzel2023} code, which incorporates physically motivated models to enhance predictive power in astrophysical modeling. 
FIRE-2 focuses on resolving the multi-phase interstellar medium (ISM), capturing the critical physical processes that govern star formation within galactic disks \citep{hopkins_2018_fire2}. 
These simulations utilize stellar population synthesis models \citep{Leitherer1999, Kroupa2001} and explicitly account for energy, momentum, and mass returned from stars.

\begin{figure*}[htbp]
\centering
\includegraphics[width=1 \textwidth]{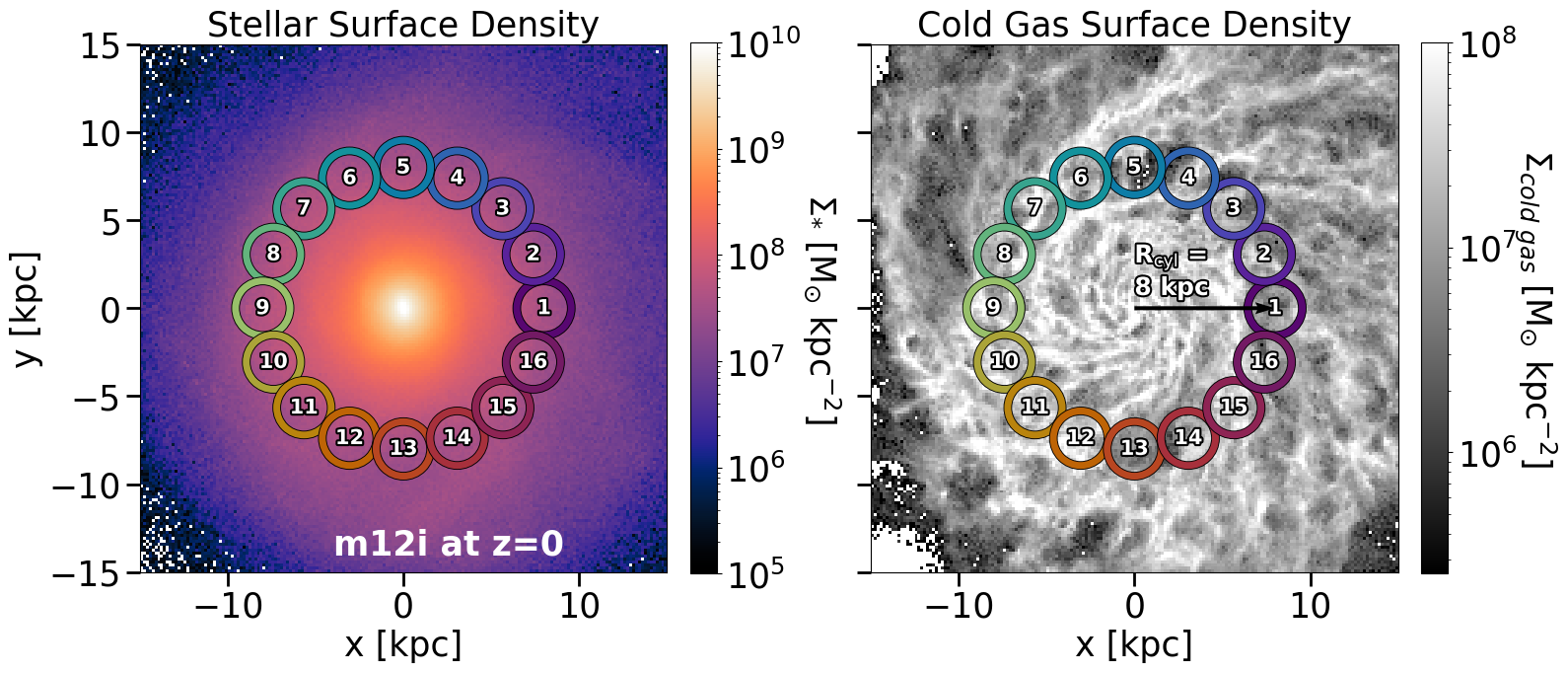}
\caption{Selecting volumes of stars across the disk and analyzing within individual volumes enables us to explore how diverse disk environments impact OTI inferences. \textit{Left:} The background displays stellar surface mass density for FIRE-2 simulated galaxy, m12i, at present-day. In the foreground, we plot open, colored circles indicating 16 solar-analog volumes, each centered at a galactocentric radius of 8 kpc, with a radius of 1.4 kpc and a height constraint of \(|z|\) \(\leq\) 3.5 kpc, containing the stellar populations analyzed in our OTI study. \textit{Right:} Equivalent plot for cold (\( T < 10^4 \text{ K} \)), dense ($n$ \( > 10 \) cm$^{-3}$) gas, revealing localized variations in the cold, dense gas distribution.
\label{fig:Figure1}}
\end{figure*}

These simulations use the Lagrangian Mesh-free Finite Mass (MFM) hydrodynamics method implemented in the GIZMO gravity plus hydrodynamics code \citep{hopkins_2015_mesh_free_hydrodynamic_simulations}, providing adaptive spatial resolution, as discussed further below. FIRE-2 encompasses a comprehensive array of physical processes and feedback mechanisms, including stellar feedback sources such as radiation pressure, stellar winds from young and old stars, photoionization, and photoelectric heating. FIRE-2 accounts for supernovae explosions with rates based on \textsc{starburst99} \citep{Leitherer1999} and the delay time distribution of type Ia supernovae from \citet{Mannucci2006}. These introduce significant energy into the ISM, generating hot gas bubbles that exert substantial pressure, resulting in the expulsion of material into the intergalactic medium \citep{FG13, Gurvich20, Orr22}. 

m12i includes stellar feedback, fluid dynamics, and star formation. A full description of the criteria used to generate m12i and its properties at present day can be found in \citet{bellardini_3d_elemental_abundances}, \citet{McCluskey24}, \citet{Bhattarai2024}, and references therein. Specifically, m12i has a total mass within the spherical radius containing 200 times the mean matter density of the universe, $M_{200} = 1.18 \times 10^{12} M_{\odot}$ \citep{Wetzel2016, Wetzel2023}. This system has a stellar mass of 6.3 $\times 10^{10} M_{\odot}$ enclosed within a spherical radius containing 90$\%$ of the stellar mass that is within 20 kpc. The initial mass of DM particles is 3.5 $\times 10^{4}$ $M_{\odot}$ and the initial mass of baryons (stars and gas) is 7070 $M_{\odot}$. Collisionless particles represent DM and stars, with a force resolution of 4 pc for stars and 40 pc for DM, while gas is represented by spatially-adaptive cells, which resolve down to a minimum spatial scale of 1 pc in the highest density regions. Star formation can proceed when gas is self-gravitating, cold (T $< 10^4$ K), dense ($n > 10^3$ cm$^{-3}$), and molecular. 

FIRE-2 datasets include full 6D kinematic phase space and detailed chemical abundances for 11 elements. In this study we consider hydrogen (H) and iron (Fe) content in star particles. Metallicity and element abundances are scaled to solar values \citep{asplund_2009_chemical_composition_of_sun}. The metallicity distribution has been well-studied in m12i, including within the spiral arms \citep{Bellardini2021, bellardini_3d_elemental_abundances, Orr23, Parul25}. Additionally, this live disk captures real-time evolutionary processes, enabling the representation of complex interactions, ranging from merger events to tidal interactions, fundamental for assessing disequilibrium effects \citep{Bonaca17, Necib2019,  Garavito-Camargo24, Arora2024c, Ou25}. 

As OTI is a modeling technique reliant on aggregate quantities used to make statistical inferences, it thus necessitates a densely-sampled phase space. The particle accelerations used for comparison with the OTI inferences are a stored quantity for \textit{Latte} galaxies. The volume of simulated data, encompassing over a million particles within the galactic disk in the FIRE-2 simulations, provides a robust foundation for assessing how disequilibrium effects influence the OTI-derived galactic mass distribution estimates.

Figure~\ref{fig:Figure1} depicts m12i top-down at present day. 
In both panels, we use bin sizes of 28 pc$^{2}$. The left panel is colored by stellar surface mass density on a logarithmic scale. We overplot 16  localized solar-analog volumes all centered on \(R=8 \, \text{kpc}\)\footnote{We have adopted fixed radius of $R=8 \, \text{kpc}$ following \citet{Arora2024}, which found this volume to match the local density of the MW. Additionally, \citet{Bellardini2021} showed that measuring FIRE-2 galaxies in physical units rather than scaling by scale length leads to less scatter and more consistency, supporting the choice of a fixed radius.}, with a radius of 1.4 kpc, spaced 22.5$^{\circ}$ apart, enabling maximal spatial coverage without overlap. We constrain heights to within \(z \pm  3.5 \, \text{kpc}\) in order to select for stars that are on more disk-like orbits, but enabling maximal disk coverage.  
We have opted not to cut on the guiding-center radius \citep[as was done in][]{pricewhelan2024datadriven, Horta2024} since we have the instantaneous accelerations which correspond to the instantaneous $R_{\mathrm{cyl}}$ for stars in the volume. The colored rings differentiate between the various solar volumes. We are sufficiently well-sampled to carry out OTI analysis, given the star particle counts per volume range from 3.4 $\times 10^{4}$ to 4.7 $\times 10^{4}$. We  analyze the stellar data that lies within each volume individually, enabling a comparison of the results between volumes. The right panel shows the azimuthal variation in surface density of cold (\(T < 10^4 \, \text{K}\)), dense (density \(> 10\) particles per cubic centimeter) gas \citep[see fractional variation in Figure 1 in][]{Arora2024}; solar volumes trace out areas of differing conditions across the entire galactic disk.

\begin{figure}
\centering
\includegraphics[width=0.48\textwidth]{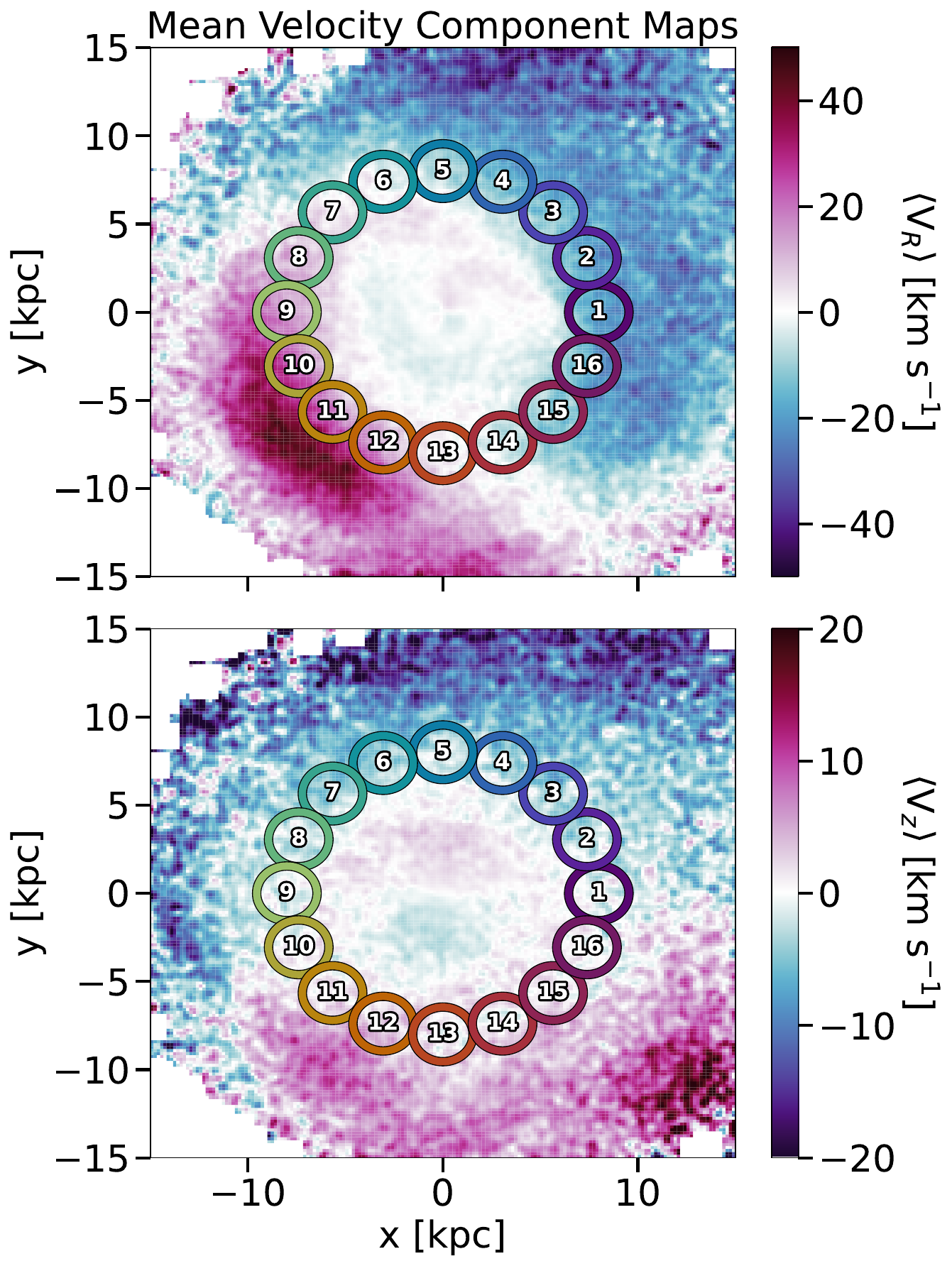}
\caption{Asymmetric global stellar kinematic conditions are present at present-day in m12i: violet and blue lobes near the solar annulus emphasize variations in stellar velocities. \textit{Top:} A top-down view of the mean stellar radial velocity. Colored circles mark the solar regions shown in Figure~\ref{fig:Figure1}. \textit{Bottom:} Similar view of the mean stellar vertical velocity. The asymmetry in velocity distributions provides insights into the kinematic structure across the galactic disk.
\label{fig:Figure2}}
\end{figure}

As this study is built on stellar chemodynamical data, we consider the global stellar motion.  Figure~\ref{fig:Figure2} shows 2D histograms of the mean stellar radial velocity \textit{(top)} and the mean stellar vertical velocity \textit{(bottom)} using bins of size 28 pc$^{2}$, as in Figure~\ref{fig:Figure1}. Stars are drawn from the disk with the same height constraint (\(z \pm  3.5 \, \text{kpc}\)) as applied in Figure~\ref{fig:Figure1}. We overplot the same 16 solar volumes from Figure~\ref{fig:Figure1}. We note asymmetrical conditions present within the solar volumes for both mean radial and vertical stellar motion, as denoted by the predominantly violet and blue lobes which can be seen near the solar annulus, indicative of a system out of steady-state. We note differing symmetries of the peaks of the distribution between the top and bottom panels, indicating the complexity of a dynamically-evolving system.

\begin{figure}
\centering
\includegraphics[width=.48\textwidth]{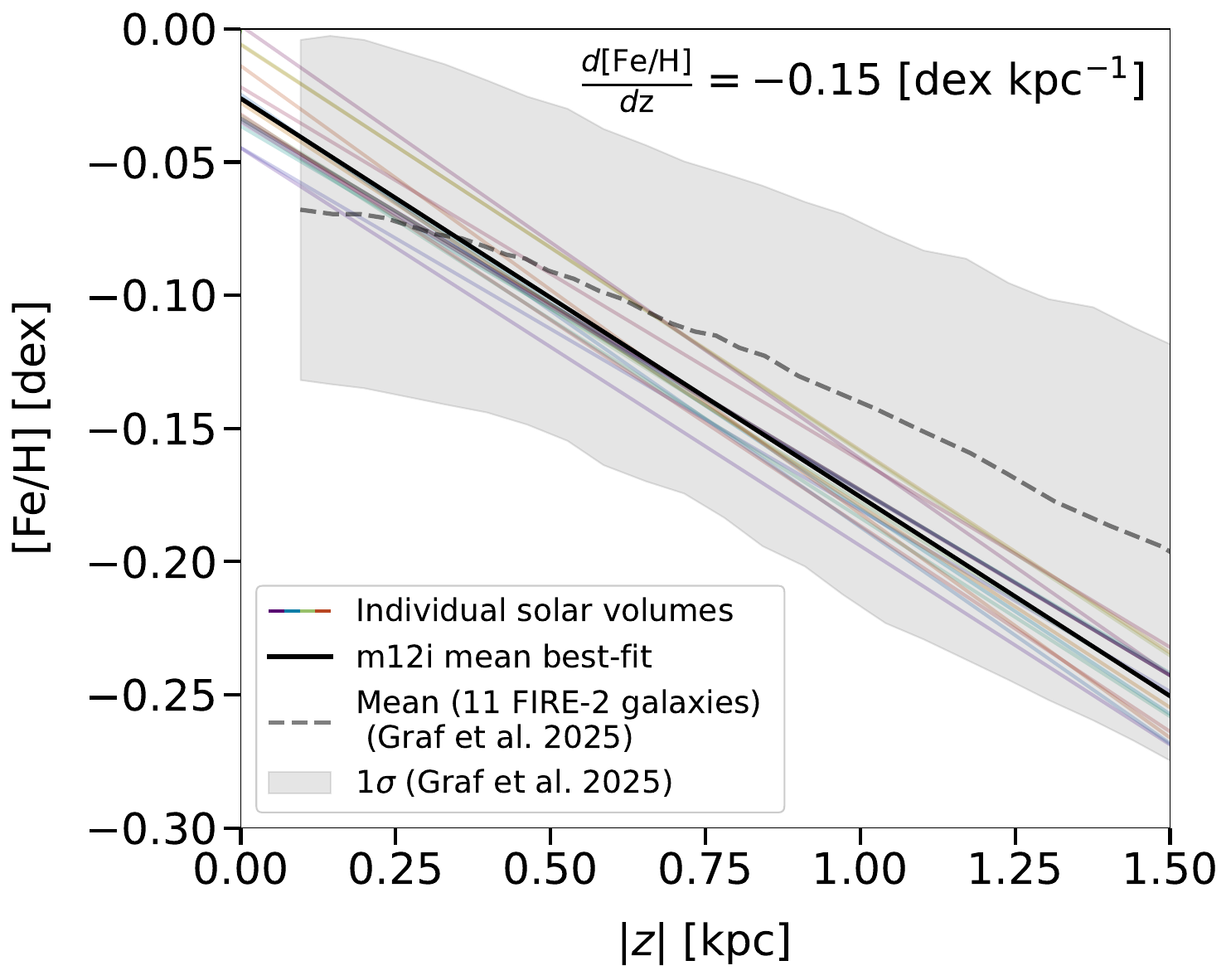}
\caption{The vertical metallicity gradient slope in  m12i at present-day is non-zero in all solar volumes, a requirement for OTI analysis. We plot the mean line of best fit across 16 solar-analog volumes (black solid line), showing a slope of \(\bar{m}=-0.15 \) \(\text{dex kpc}^{-1}\), as well as the best-fit line in all 16 solar volumes (colored translucent lines). See Figure~\ref{fig:feh_gradient} in Appendix~\ref{sec:appendixa} for the measured slopes of each individual volume. For reference, we include the black dashed line representing the mean vertical [Fe/H] profile across all stars in 11 FIRE-2 MW-mass galaxies at present day at $R$ = 8 kpc, and the 1\(\sigma\) interval (black shaded region) \citep{Graf2025}. For comparison, the MW has vertical metallicity gradient slope estimates from -0.1 to -0.3 \(\text{dex kpc}^{-1}\) \citep{Ivezic2008, Carrell2012, Hayden2014, Imig2023}. See \S\ref{sec:simulations} for further discussion.
\label{fig:Figure3}}
\end{figure}

\begin{figure}
\centering
\includegraphics[width=.48\textwidth]{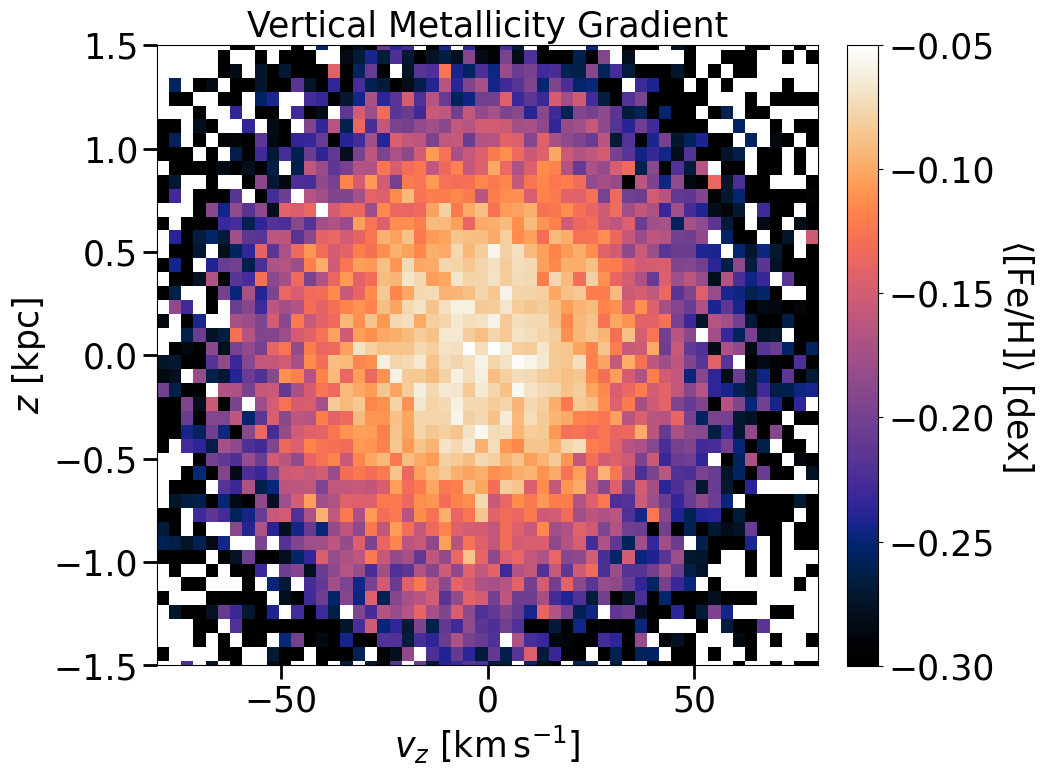}
\caption{The average mean [Fe/H] across all 16 volumes in bins of $z$-$v_{z}$. Mean [Fe/H] systematically decreases from central to outer regions, as seen in the MW; this strong vertical metallicity gradient improves the OTI fit for parameterized orbital shapes. Such a gradient enables accurate orbital modeling, as it maps more distinct orbital paths to the vertical distribution. See \S\ref{sec:simulations} for further discussion and Figure~\ref{fig:FigureA} in Appendix \ref{sec:appendixa} for individual vertical metallicity gradients within each volume.
\label{fig:Figure4}}
\end{figure}

We additionally consider how metallicity changes across m12i at present-day, as this is a crucial component in OTI modeling. Particularly, we are interested in ensuring that m12i replicates the observed trend of differing mean element abundance value with varying galactic heights, similar to the vertical metallicity gradient seen in the MW \citep{APW21, Horta2024}.

We confirm this trend is present across all volumes (see Appendix \ref{sec:appendixa} for the slope plotted for each volume individually). Figure~\ref{fig:Figure3} shows the mean line of best-fit to the stellar [Fe/H] data across all solar volumes in m12i (black solid line), and best-fit lines within each volume (colored translucent lines). We compute the best-fit lines using a least squares regression and find the mean best-fit slope is  $\bar{m}=-0.15$ \(\text{dex kpc}^{-1}\). For reference, we also include results from \cite{Graf2025} which show the mean vertical [Fe/H] profile (black dashed line) for all stars within 11 FIRE-2 MW-mass galaxies and the corresponding 1\(\sigma\) interval (black shaded region). As in \cite{Graf2025}, we show metallicity profiles at present day at $R=8$ kpc. We find our results are consistent with these other FIRE-2 galaxies.
Estimates for the MW's vertical metallicity gradient slope range from -0.1 to -0.3 \(\text{dex kpc}^{-1}\) \citep{Ivezic2008, Carrell2012, Hayden2014, Imig2023}, steeper than the FIRE-2 slopes we present. \citet{Graf2025} suggests this discrepancy may stem from the MW's disk settling earlier than those in FIRE-2 or experiencing less mixing, both of which would preserve a stronger vertical metallicity gradient. While the slopes of the vertical metallicity gradients we present are shallower, they are measurable and qualitatively consistent with the MW's, making them suitable for OTI analysis.   

As in Figure~\ref{fig:Figure3}, we confirm that the vertical metallicity gradient for FIRE-2 simulated stellar data is consistent with that observed for mean [Fe/H] in the MW \citep[see Figure 2 from both][]{APW21, Horta2024}. Figure~\ref{fig:Figure4} shows an azimuthal average of $\langle$[Fe/H]$\rangle$. Bin sizes range from 170 to 226 (pc $\times$ km s$^{-1}$) due to the dispersion within volumes. Across the entire galactic disk we note a smooth transition from higher values (yellow) of $\langle$[Fe/H]$\rangle$ (corresponding to lower absolute values of $z$ and $v_z$) to lower values (purple) of $\langle$[Fe/H]$\rangle$ (corresponding to higher absolute values of $z$ and $v_z$). This is the same smooth, systematic variation as observed in \citet{APW21}, using APOGEE and Gaia data. This vertical metallicity gradient is what the OTI modeling leverages to infer potential parameters. By corresponding the level set contours to orbits and parameterizing those orbital shapes, we are able to estimate galactic potential parameters \citep[see \S\ref{sec:methods} and][for further description]{pricewhelan2024datadriven}. While we report a measurable vertical metallicity gradient for the other element abundance ratios tracked in FIRE-2, such as $\langle$[Mg/Fe]$\rangle$, $\langle$[Fe/H]$\rangle$ was the steepest, motivating its use in OTI analysis.

\section{Orbital Torus Imaging} \label{sec:methods}
The method introduced by \citet{APW21} is based on the concept that, in steady-state, stellar invariants -- like age or element abundance -- are functions solely of dynamical quantities -- like energy or angular momentum -- and do not have an explicit dependence on orbital phase. 
This is the primary idea that underlies the OTI framework in that, by using the contours of constant label along an orbit, which are manifest in the vertical phase space ($z-v_{z}$) distribution, it is possible to trace out the orbit structure of stars in that system. 
We empirically observe smooth gradients of element abundance ratios in vertical phase space in the MW; OTI leverages these gradients to infer potential parameters \citep{APW21, Horta2024}. 

The updated implementation from \citet{pricewhelan2024datadriven} fits a flexible model describing orbit shape to the level set contours of the mean element abundance ratio ([X/Y]) in bins of $z-v_{z}$. From there, it optimizes the likelihood of the model parameters given the mean element abundance distribution ($\langle$[X/Y]$\rangle$($z$, $v_{z}$)), and, from this best-fit model, derives quantities like the vertical acceleration. The full description of model parameters and the OTI methodology is outlined in \citet{pricewhelan2024datadriven}.

The assumptions included in the OTI framework are equilibrium, phase-mixing, an axisymmetric system, and separability of $R$ and $z$ variables -- a reasonable assumption for near-circular orbits \citep{BT2008}. We employ the the 1D vertical collisionless Boltzmann equation \citep[Equation 11 from][]{pricewhelan2024datadriven}, considering only the vertical kinematics.\footnote{We neglect the radial term since the radial bins are sufficiently small that the radial gradient is flat. 
We note that this approximation may not be valid for regions far ($\left| z \right|$ $\geq$ 1.5 kpc) from the midplane of the galactic disk, however, we restrict our analysis region closer to the galactic midplane, where the assumption of $R$-$z$ separability is more valid.} 
OTI models orbital shapes as a function of the radius in the $z-v_{z}$ plane. A proxy vertical action, $r_{z}$, represents a Fourier distortion away from an elliptical shape. It is defined in terms of the elliptical radius, $\tilde{r}_{z}$, and the elliptical angle, $\tilde{\theta}_{z}$ \citep[see Equation 21 from][]{pricewhelan2024datadriven}. This parameter captures how the orbital shape evolves as a function of the radius in the $z-v_{z}$ plane, reflecting the changing ratio of baryonic to dark matter. As stellar trajectories transition from the vicinity of the midplane to larger heights, their shapes shift from ellipses to more pinched, diamond-like orbits. OTI enables flexible representations of orbit shapes that are more complex than ideal ellipses. These changes can occur since, as a star nears the midplane, it feels a rapidly-varying gravitational potential.

We note OTI computes the vertical acceleration according to Equation 18 from \citet{pricewhelan2024datadriven}, which we then directly compare to the equivalent quantity stored in FIRE-2 simulation m12i (see \S\ref{sec:application} for further discussion).

\section{OTI Application on m12i} 
\label{sec:application}

\subsection{Pre-Processing FIRE-2 m12i Data}
\label{pre-processing}
As we describe in \S\ref{sec:simulations}, we analyze 16 localized solar-analog volumes as it is crucial to work with regions small enough to preserve the detailed features of the measured vertical acceleration profile. This profile is highly sensitive to the local density, and using larger regions would risk averaging over areas with significantly different vertical properties (e.g., higher surface mass density at smaller $R$ and lower density at larger $R$).

The OTI model is fit to \(\langle\)[Fe/H]\(\rangle\), the mean metallicity, in bins of $z-v_{z}$. Within each solar volume, we bin the star particle data using 100 pixels spanning (-$z_{\rm{max}}$, $z_{\rm{max}}$) and (-$v_{z, \rm{max}}$, $v_{z, \rm{max}}$) within each volume, computing \(\langle\)[Fe/H]\(\rangle\) in each bin. The median absolute deviation (MAD) is a measure of statistical dispersion \citep{Beers90}.
It is a robust estimator for $\sigma$ such that 1.4826 $\times$ MAD $\sim$ $\sigma$.
In order to capture 3$\sigma$ of our distribution, $z_{\rm{max}}$ and $v_{z, \rm{max}}$ are defined as 3 $\times$ 1.5 $\times$ MAD($z$) or MAD($v_{z}$), respectively. Due to the dispersion within each volume, $z_{\rm{max}}$ ranges from (3.0, 3.7) kpc and $v_{z, \rm{max}}$ ranges from (139, 154) km s$^{-1}$. Following \citet{pricewhelan2024datadriven}, we infer the intrinsic scatter based on all bins containing at least 32 star particles. OTI can work with any moments of the stellar label (abundance) distribution, but here we stick to using \(\langle\)[Fe/H]\(\rangle\) to be analogous to \citet{pricewhelan2024datadriven}. These are the relevant features used in OTI analysis, shown across the entire disk in Figure~\ref{fig:Figure4} for \(\langle\)[Fe/H]\(\rangle\). We bin the data since OTI relies on aggregate quantities to make statistical inferences from a well-sampled phase space. As described in \S\ref{sec:application}, OTI does this by analyzing orbital shapes set by the mean element abundance distribution in $z-v_{z}$, or any property that maps out the contours of the distribution function.

\begin{figure*}[htbp]
\centering
\includegraphics[width=1\textwidth]{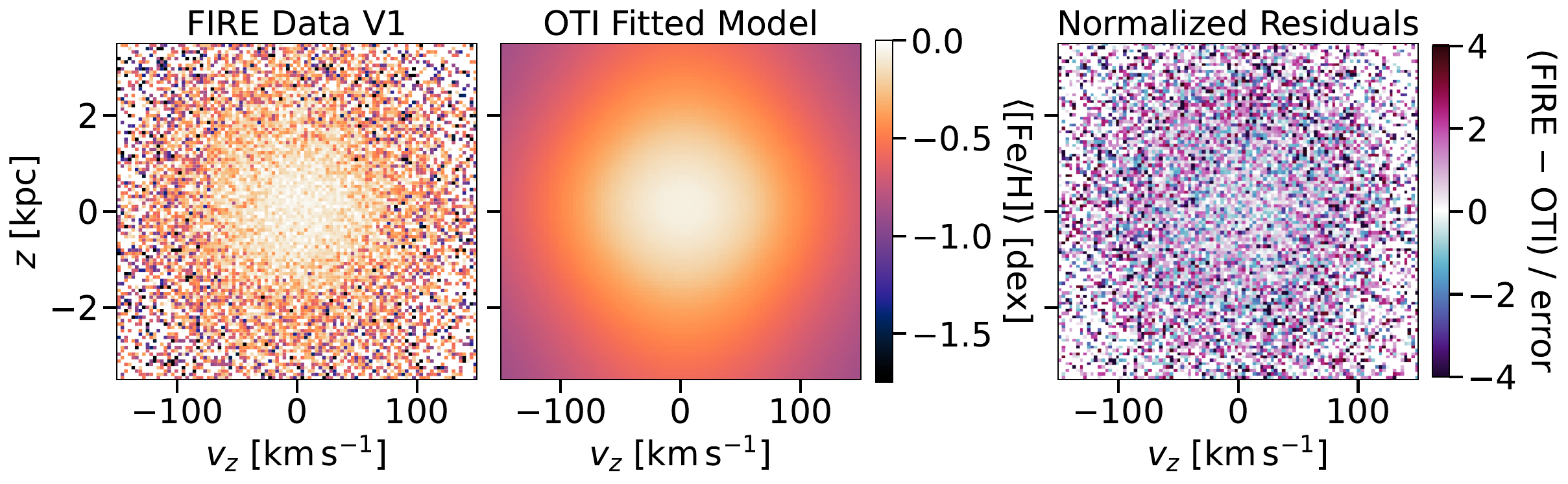}
\caption{The OTI model reliably captures the known FIRE-2 \(\langle\)[Fe/H]\(\rangle\) distribution with nonsystematic residuals. \textit{Left:} 2D histogram of the \(\langle\)[Fe/H]\(\rangle\) abundance from Volume 1 (V1) in m12i at present-day, in bins ranging from 170 to 226 (pc * km s$^{-1}$), as in Figure~\ref{fig:Figure4} but for a single volume. \textit{Middle:} OTI-optimized model. \textit{Right:} Residual plot of FIRE $-$ OTI mean [Fe/H], normalized by the intrinsic scatter within each pixel. See \S\ref{sec:optimization} for further discussion and Appendix \ref{sec:appendixa} for the vertical metallicity gradients within each volume.
\label{fig:Figure5}}
\end{figure*}

\subsection{OTI Model Optimization}
\label{sec:optimization}
We use the \texttt{torusimaging} package \citep{pricewhelan2024datadriven} to fit an OTI model to FIRE-2 mean metallicity, binned in $z-v_{z}$, for each of our solar-analog volumes. 
Using \texttt{torusimaging}, we model the dependence of \(\langle\)[Fe/H]\(\rangle\) as a function of $r_z$  in vertical phase-space by parameterizing the shapes of contours in the mean abundance distribution.
The free parameters of our OTI model describe flexible functions of the shape and spacing of these contours.
Following \citet{pricewhelan2024datadriven}, we use an OTI model with both $m=2$ and $m=4$ Fourier distortions, and model parameters that define spline knot values for: 

\begin{enumerate}
    \item The dependence of \(\langle\)[Fe/H]\(\rangle\) on the proxy vertical action, $r_z$,
    \item $e_2(\tilde{r}_z)$, the $m=2$ distortion function,
    \item $e_4(\tilde{r}_z)$, the $m=4$ distortion function.
\end{enumerate}

OTI uses a monotonic quadratic spline to represent the orbital shape, and the \textit{knots} are the points where two polynomial segments are joined together to form a smooth curve. Exact hyperparameters used can be found in Table \ref{tab:initial_model_properties} in Appendix \ref{sec:appendixc}.
We use 8 spline knots for the label function, 10 knots for the $e_{2}$ function, and 5 knots for the $e_{4}$ function, along with fitting for the centroid of the mean abundance distribution in phase space $(z_0, v_{z,0})$, and the asymptotic midplane value of the orbital frequency, $\Omega_0$.
In total, our OTI model has 26 free parameters.

OTI employs JAX (Just After eXecution) \citep{bradbury2018}, a machine/deep learning Python library for high-performance numerical computing, to automatically differentiate the objective function with respect to our model parameters. This efficient gradient computation enables informed updates to the model parameters as the optimizer fits the model for orbital shapes to the input data, here contours of average stellar metallicity. 

We use the L-BFGS optimization scheme \citep{Byrd1995}, a quasi-Newton method, to fit the model to the data. Following \citet{pricewhelan2024datadriven}, we use a Gaussian log-likelihood for the \(\langle\)[Fe/H]\(\rangle\) data in each $(z, v_z)$ bin (see Equation 33). We provide initial parameters based on the binned FIRE-2 \(\langle\)[Fe/H]\(\rangle\) values and set bounds on these parameters according to the sign of the first derivative of the quadratic spline at the knot locations; we enforce monotonicity since $J_{z}$ varies smoothly with $r_{z}$. We optimize after masking for only finite values to obtain a parameterized best-fit OTI model to FIRE-2 data, from which $a_{z}$ can be computed \citep[see Equation 18 from][]{pricewhelan2024datadriven}. 

Figure~\ref{fig:Figure5} displays such an OTI fitted model for just one solar volume from m12i at present-day (solar volume 1), as an example. The left panel shows our input data: FIRE-2 \(\langle\)[Fe/H]\(\rangle\) binned in $(z, v_z)$ with sizes ranging from 170 to 226 (pc $\times$ km s$^{-1}$), as in Figure~\ref{fig:Figure4}.
The middle panel shows the optimized OTI model, displaying the model-predicted \(\langle\)[Fe/H]\(\rangle\) in the same bins as the left panel.
The right panel shows the residuals between the FIRE-2 data and the best-fit OTI model (left panel minus middle panel), normalized by the intrinsic scatter of the abundances within each bin.

In each volume we note a smooth transition from higher values (yellow) of \(\langle\)[Fe/H]\(\rangle\) (corresponding to lower absolute values of $z$ and $v_z$) to lower values (purple) of \(\langle\)[Fe/H]\(\rangle\) (corresponding to higher absolute values of $z$ and $v_z$), also seen in the MW \citep{APW21, Horta2024}. Note the visual similarity between the OTI fitted model and the true FIRE-2 metallicity gradient. While there is a faint ring of darker pixels--an expected effect of shot noise due to lower particle counts in those bins--the normalized residuals overall appear uniformly distributed, indicating nonsystematic trends in the residuals. The accuracy in capturing the known FIRE-2 vertical metallicity gradient consequently enables an accurate prediction of the known FIRE-2 potential parameters. 

\section{Results from OTI application to m12i} \label{sec:results}

\begin{figure*}[h]
\centering
\includegraphics[width=1\textwidth]{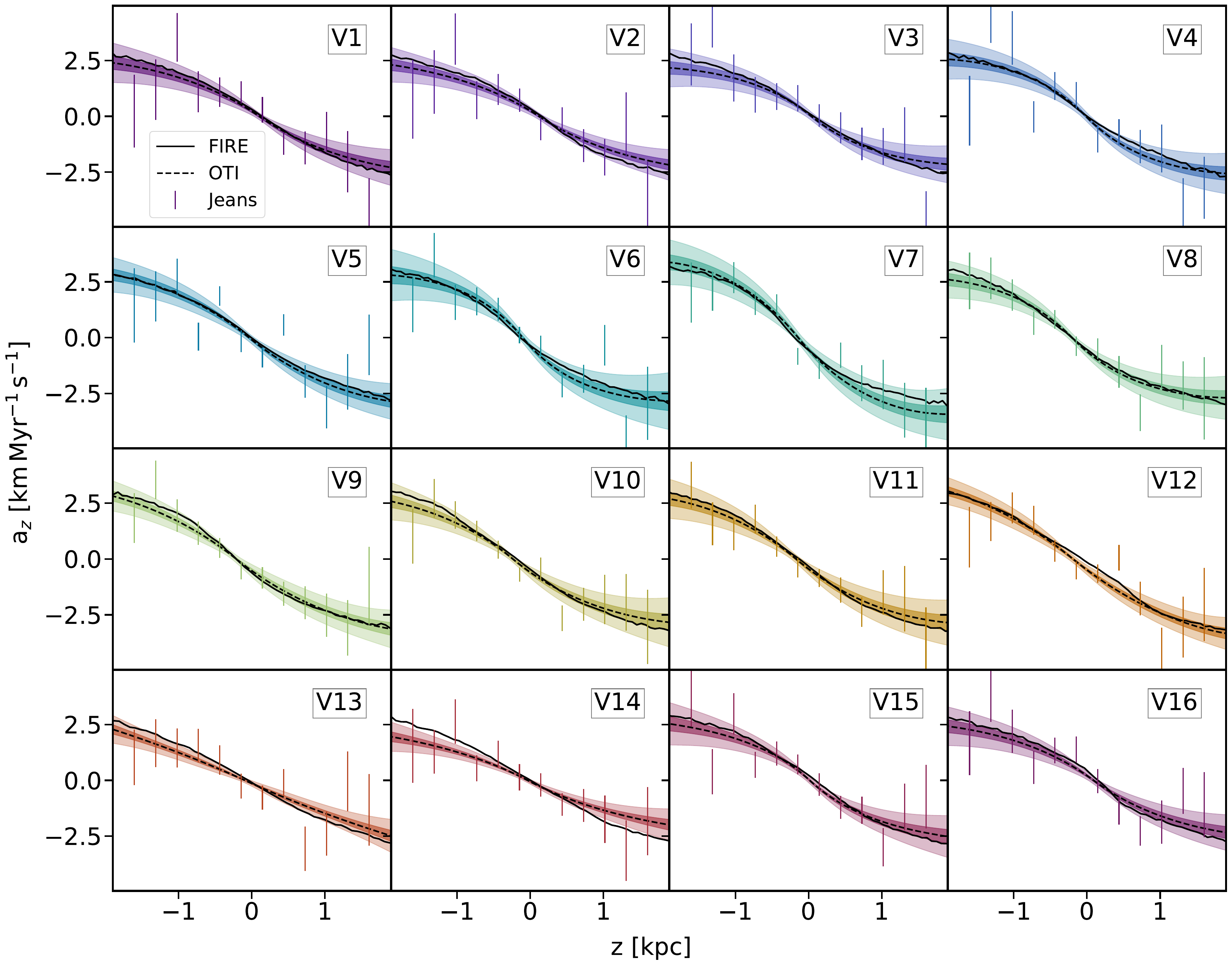}
\caption{The OTI model accurately recovers the true vertical acceleration profile within 3\(\sigma\) for 15 out of 16 solar volumes (all but V14) and within 1\(\sigma\) for 12 out of 16 solar volumes (all but V2, V13, V14, and V16) in m12i. The black solid line represents FIRE-2 simulated galaxy m12i's vertical acceleration profile, while the black dashed line represents the OTI model inference. The darker shaded region indicates the 1\(\sigma\) uncertainty, and the lighter shaded region shows the 3\(\sigma\) limit. We compute the uncertainty by MCMC sampling the posterior distribution, bootstrapping the FIRE-2 data, and combining the variances in quadrature to determine the total error. Vertical solid lines show the Jeans modeling $a_{z}$ estimates with errors calculated from bootstrapping with replacement. We discuss several properties of the local disk that are correlated with a decrease in OTI accuracy in \S\ref{sec:disc} and in Appendix~\ref{sec:appendixg}.
\label{fig:Figure6}}
\end{figure*}

\subsection{a$_{z}$ Profiles}
\label{sec:a_z}
We use the OTI model and setup outlined in \S\ref{sec:methods} to infer vertical acceleration, $a_{z}$, profiles for all 16 solar volumes. Figure~\ref{fig:Figure6} shows the true FIRE-2 vertical acceleration profile (black solid line) as compared with the OTI-inferred vertical acceleration profile (black dashed line). We compute the uncertainty of the inferred acceleration profile using two approaches: one, by applying priors to our parameters and performing MCMC sampling, and two, by repeatedly optimizing the model after bootstrap resampling of the data. We initialize MCMC chains using parameter values we obtain from optimization (see \S\ref{sec:optimization} for further details). We combine the MCMC and bootstrapping variances in quadrature to arrive at a total (MCMC + bootstrapping) error. The darker shaded region is 1$\sigma$ of the total uncertainty and the lighter shaded region is 3$\sigma$ of the total uncertainty. We use the No-U-Turn Sampler implemented in the \texttt{blackjax} package \citep{Cabezas2024} to explore the posterior distribution of our model parameters, yielding maximum a posteriori values that represent the most probable vertical acceleration profiles. To bootstrap the data, we initialize a random number generator, conduct 128 resampling trials with replacement on the FIRE-2 binned \(\langle\)[Fe/H]\(\rangle\) data, and optimize the OTI model on each resampled dataset, storing the results for further analysis.

Figure~\ref{fig:Figure6} demonstrates a general agreement between the OTI-inferred $a_{z}$ profiles and the true FIRE-2 $a_{z}$ profiles, agreeing within 1$\sigma$ for 75$\%$ of the volumes. There are four cases that lay outside of the 1$\sigma$ range except for at the midplane: V2, V13, V14, and V16. While we consider these volumes relatively inaccurate, V2, V13, and V16 agree within the 3$\sigma$ limit, while V14 does not. We summarize several local disk properties that contribute to this loss in accuracy in \S\ref{sec:disc} and discuss these in further detail in Appendix~\ref{sec:appendixg}.

For comparison, we consider how an alternative method, Jeans modeling \citep{Jeans1922}, performs in each solar volume. This methodology is commonly used \citep{Garbari12} and assumes a steady-state system.
To obtain a Jeans modeling estimate, we apply the vertical Jeans equation (Equation \ref{eq:jeans}) to FIRE-2 data, ignoring cross-terms between radial and vertical velocity under the assumption of separability of $R$ \& $z$ coordinates.  
\begin{equation}
    \frac{\partial(\nu\sigma_{v_{z}}^{2})}{\partial z}+\nu\frac{\partial\Phi}{\partial z}=0
    \label{eq:jeans}
\end{equation}
Here, $\nu$ is the number density of FIRE-2 star particles and $\sigma_{v_{z}}$ is the vertical velocity dispersion of FIRE-2 star particles, obtained by binning the particle data in the vertical kinematics. We use a larger bin size (70 pc for FIRE-2 \& OTI, 290 pc for Jeans modeling) because it decreases the noise of the spline interpolation for the gradient of $\nu \sigma_{v_{z}}^{2}$. The vertical acceleration is computed according to 

\begin{equation}
    a_{z}=\frac{1}{\nu} \frac{\partial(\nu\sigma_{v_{z}}^{2})}{\partial z}
    \label{eq:az}
\end{equation}

We bootstrap the particle data with replacement using 64 trials before binning to determine the error bars associated with the Jeans modeling analysis. Across all 16 solar volumes, the Jeans modeling estimates (vertical colored lines) have broader uncertainty than OTI -- \textit{that is, for all volumes, OTI provides a more precise $a_{z}$ estimate by $\sim85\%$ across the disk}. 

Ultimately, we hope to understand how accurately the OTI model can recover the true mass distribution. This is particularly relevant in the case where Jeans modeling fails due to  disequilibrium effects (i.e.~disequilibrium caused by large-scale galactic interaction and/or local stellar feedback processes). For a detailed investigation into the systematic uncertainties associated with deriving potential parameters from the Jeans' equations, we refer the reader to \citet{Ou25}.

\begin{figure}[t!]
\centering
\includegraphics[width=0.48\textwidth]{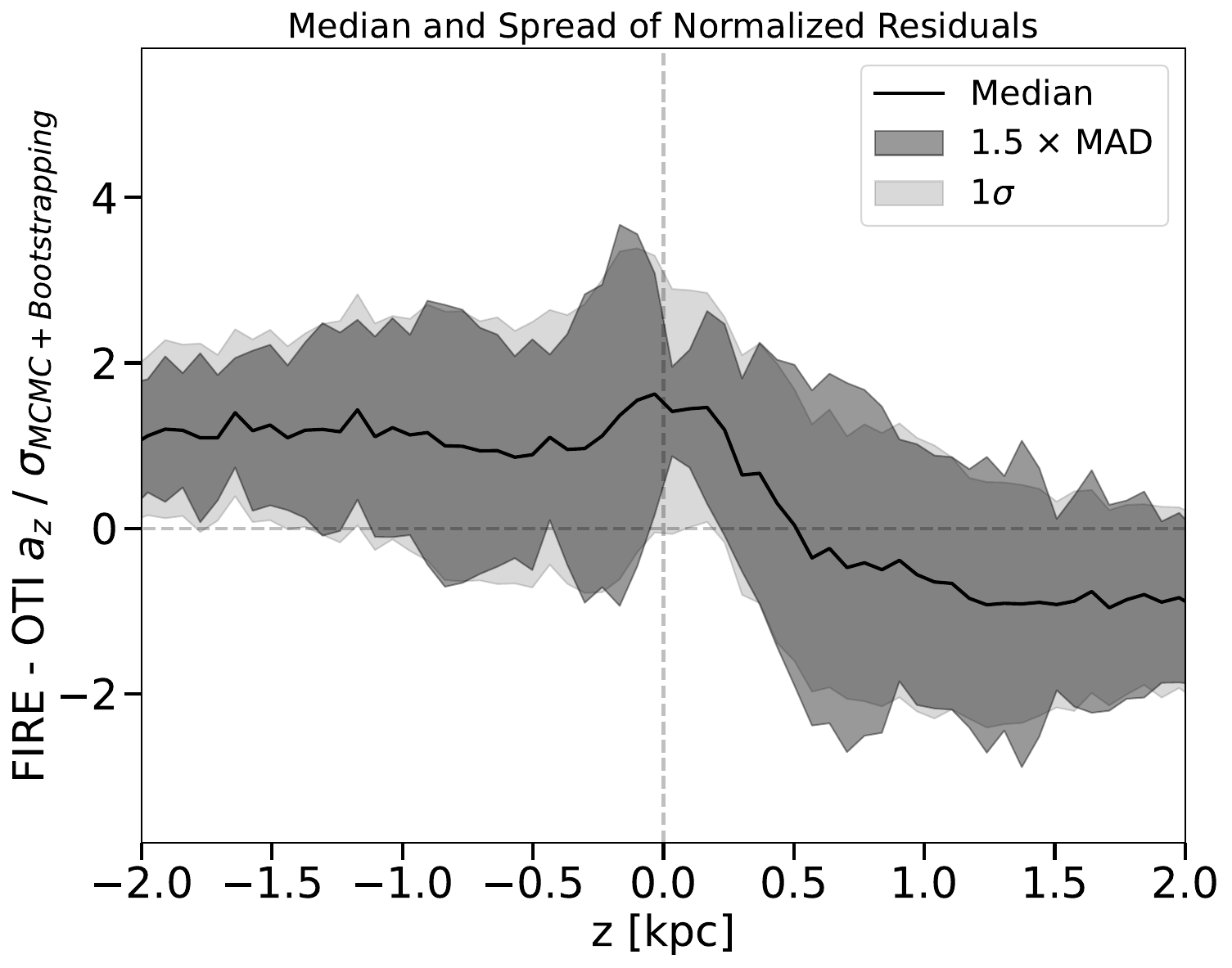}
\caption{The median (across all 16 volumes) normalized residual profile shows the OTI model achieves reasonable accuracy in recovering FIRE-2 $a_{z}$ profiles. We also plot 1$\sigma$ (light shaded region from the median, and 1.5 $\times$ MAD (darker shaded region). This provides an estimate of the model accuracy across the galactic disk, revealing a systematic offset relative to the midplane: OTI overpredicts above the midplane and underpredicts below the midplane. See \S\ref{sec:a_z} for further discussion.
\label{fig:Figure7}}
\end{figure}

In order to quantify the OTI model's match to the FIRE-2 simulated ground truth for the $a_z$ profiles, we compute normalized residuals. Figure~\ref{fig:Figure7} shows the FIRE-2 $-$ OTI $a_{z}$ residual profiles, normalized by the total uncertainty for the mean of all volumes (solid black line), with the light gray shaded region indicating 1$\sigma$ from the mean and the darker shaded region indicating 1.5 $\times$ MAD, which approximates $\sigma$ \citep[][]{Beers90}, as discussed in \S\ref{sec:application}.  The mean of the normalized residuals reflects the OTI model’s capacity to accurately predict the known FIRE-2 $a_z$ profiles across all 16 solar volumes. Overall, we see a trend of OTI underpredicting the FIRE-2 $a_{z}$ below the midplane and overpredicting the FIRE-2 $a_{z}$ above the midplane. There is an increase in the spread close to $z$=0, which is also where values for the percent difference blow up due to division by an increasingly small number. The average median percent change below the midplane is 15$\%$ and 14$\%$ above the midplane. We believe the combined uncertainty from MCMC sampling and bootstrapping the data is \textit{underestimated}, in that it reflects the precision of our measurement but not model inaccuracies. For example, OTI assumes a symmetric distribution, an assumption which is violated in a realistic simulated galaxy, m12i.  This can explain the large residual values seen in this figure, which arise from an underestimation of the errors based on the model assumptions. 

\subsection{Total Surface Mass Density}
 
\label{sec:sigma_star}
\begin{figure*}[htbp]
\centering
\includegraphics[width=1\textwidth]{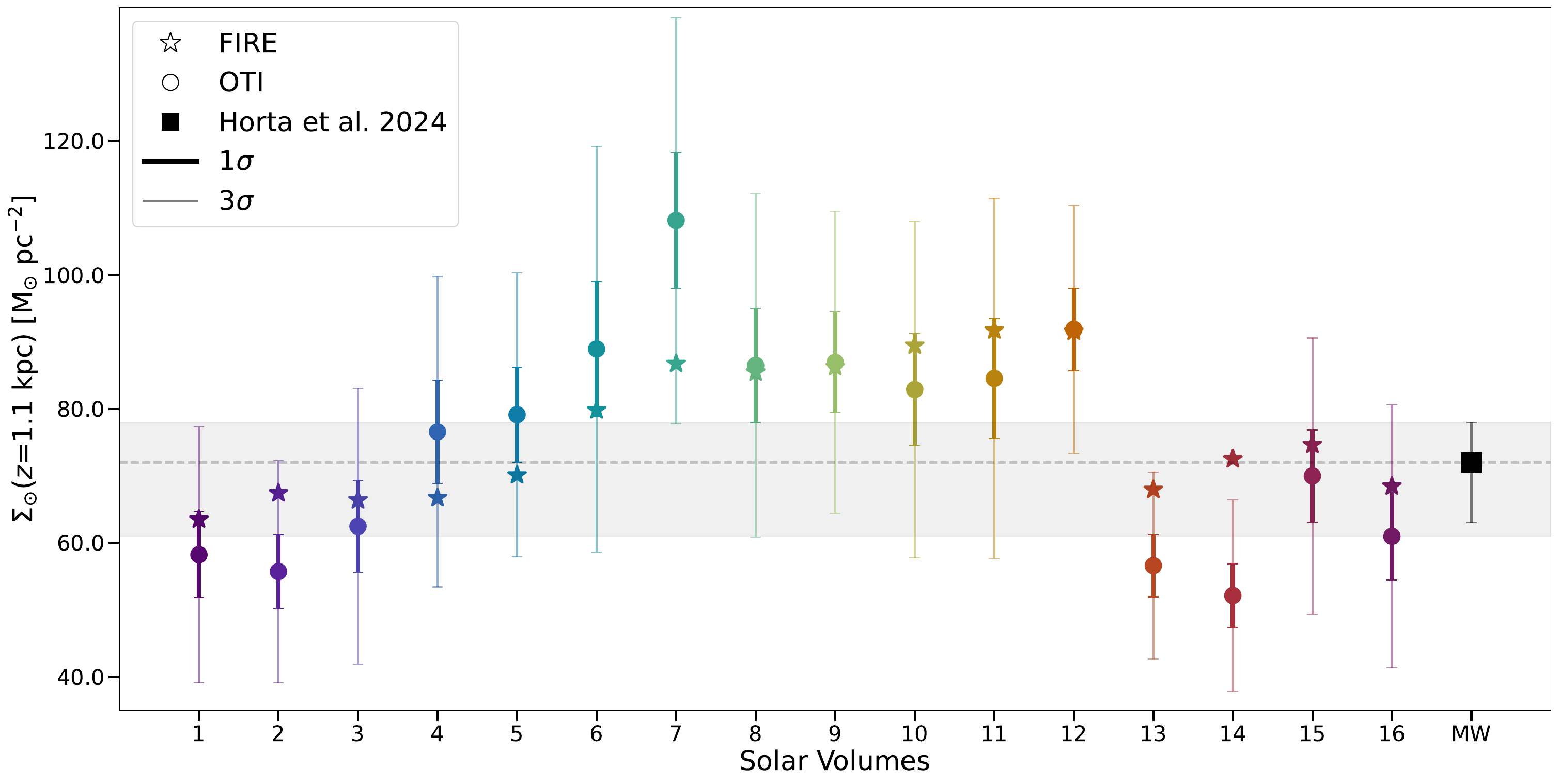}
\caption{OTI estimates agree with FIRE-2 true values for total surface mass density at $z$=1.1 kpc within 3\(\sigma\) except for V14. Total surface mass density for FIRE (star symbols) and OTI (circle symbols) is shown with 1$\sigma$ and 3\(\sigma\) error bars at $|z|$=1.1 kpc. The gray-shaded panel indicates the range of possible MW solar values from the literature \citep[Table 1,][]{Horta2024}. For comparison, we include the OTI-inferred MW total surface mass density from \citet{Horta2024}, which lies well within the range of values from our simulated solar volumes sample. See \S\ref{sec:sigma_star} for further discussion.
\label{fig:Figure8}}
\end{figure*}

Next, we consider a traditional benchmark, $\Sigma_{\odot}(z=1.1$ $\mathrm{kpc})$, which is the total surface mass density at $z$=1.1 kpc \citep{KG91}. Figure~\ref{fig:Figure8} shows this value for all volumes \citep[computed according to Equation 19 from][]{Horta2024}. Filled stars represent the FIRE-2 values, while filled circles represent the OTI inferences. OTI estimates include 1$\sigma$ (thick solid line) and 3$\sigma$ (thin solid line) error bars. The light gray shaded region indicates possible MW values found in the literature \citep[see Table 1 from][]{Horta2024}. The solid black square indicates an OTI-derived estimate for the MW's solar neighborhood using an APOGEE-Gaia crossmatched sample from \citep{Horta2024}. The OTI-predicted MW value lies squarely within the range of densities we probe, and, for all but one FIRE-2 simulated volume (V14), OTI is accurate within 3$\sigma$. In this limit, OTI is a promising new tool for constraining true MW potential parameters, like $\Sigma_{\odot}$, given its demonstrated constraining power in a simulated galaxy with comparable characteristics.

\section{Discussion} 
\label{sec:disc}

Thus far, we have applied OTI to FIRE-2 data and found that it matches the true $a_{z}$ profile to within 3$\sigma$ for 94$\%$ of cases and 1$\sigma$ for 75$\%$ of cases. Here, we examine the physical properties that are connected with OTI’s success in the majority of cases and discuss some of its limitations. In order to discuss this, we consider several aggregate statistics (scale height, asymmetry, strength of the metallicity gradient, total surface mass density, and median stellar age) 
and how they correspond with the goodness-of-fit, $\chi^{2}$, in each volume. Figure~\ref{fig:pccm} in Appendix \ref{sec:appendixf} displays these Pearson correlation coefficient matrix for these quantities.

\begin{figure*}[htbp]
\centering
\includegraphics[width=1\textwidth]{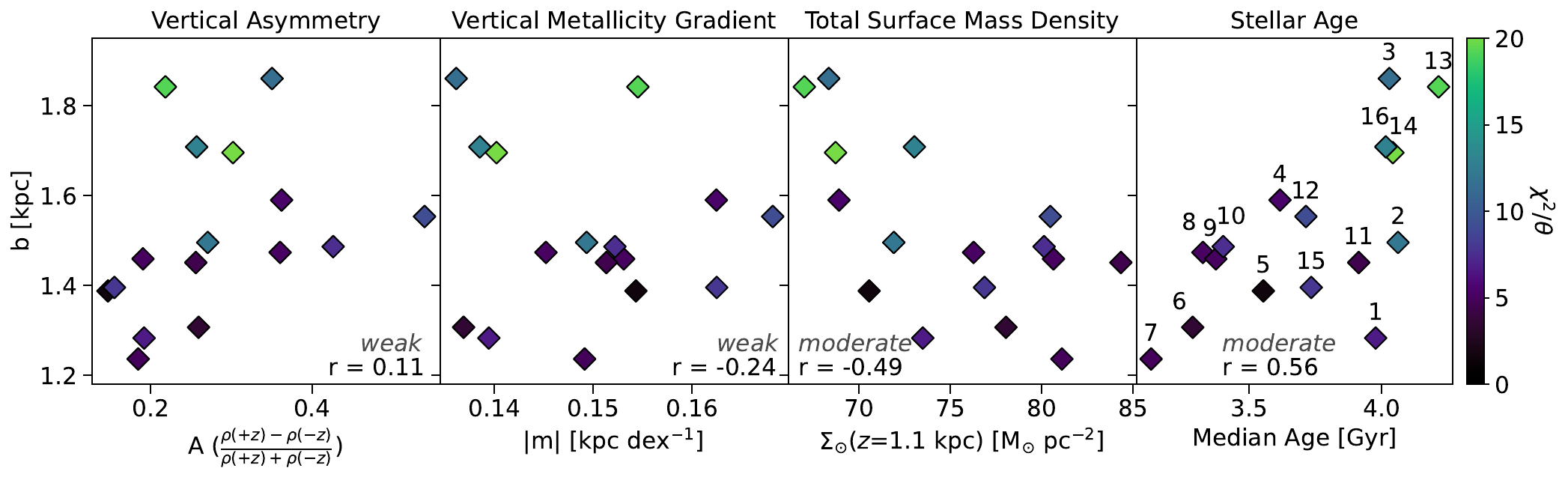}
\caption{Vertical asymmetry (\textit{left}) and slope of the metallicity gradient (\textit{middle left}) have weak correlations with the goodness-of-fit, \(\chi^{2}\), normalized by $\theta$, the total number of parameters, while total density (\textit{middle right}), median age (\textit{right}), and scale height moderately correlate with \(\chi^{2}\). Here, r refers to the Pearson correlation coefficient between each quantity plotted on the horizontal axis and \(\chi^{2}\) (see Figure~\ref{fig:pccm} in Appendix \ref{sec:appendixf} for r between all quantities). OTI can reasonably accommodate a range of vertically asymmetric distributions and a range of slopes, but results in a loss in accuracy for volumes where the total density is low, the median age is high, or the scale height is high. See \S\ref{sec:disc} and Appendix~\ref{sec:appendixg} for further discussion.
\label{fig:Figure9}}
\end{figure*}

We calculate the goodness-of-fit according to 

\begin{equation}
    \chi^{2}= \sum_{i}\frac{(O_{i}-C_{i})^{2}}{\sigma_{i}^{2}} 
    \label{eq:chi}
\end{equation}

\noindent where $O_{i}$ are the true FIRE $a_{z}$ values, $C_{i}$ are the calculated OTI $a_{z}$ values, and $\sigma_{i}$ are the OTI model's MCMC and bootstrapping uncertainty. In Figure~\ref{fig:Figure9}, we color-code our data points by $\chi^{2}$, with dark colors indicating a good match and light colors indicating a poor match.

One of the aggregate properties we consider in Figure~\ref{fig:Figure9} is the scale height. To calculate the scale height we fit a single Miyamoto-Nagai (MN) profile to the vertical stellar density at a fixed radius of $R=8 \, \text{kpc}$, using a custom model template from the \texttt{astropy} package \citep{Astropy18}. This quantity is shown on the vertical axis of all panels in Figure~\ref{fig:Figure9}. We note there are a range of scale heights across our sample from scale heights as high as 1.9 kpc and as low as 1.3 kpc. 

On the leftmost panel of Figure~\ref{fig:Figure9}, we show scale height of each volume plotted against the asymmetry of the distribution about the midplane (set as the peak of the density distribution within each volume). We compute the asymmetry by subtracting the total density below the midplane from the total density above the midplane following the prescription described in Equation 8 from \citet{Bennett2022}, but for the volume density instead of the number density. The vertical density profiles for stars, gas, DM, and total distribution within each volume can be seen in Figure~\ref{fig:vertical_density_profiles} in Appendix \ref{sec:appendixb}. Figure~\ref{fig:smd_2dhist} displays a side-on view of the stellar distribution in $R$ and $z$, providing a global perspective on warping within and around each volume.
Figure~\ref{fig:Figure9} shows there is almost no correlation (Pearson's r=0.11) between the asymmetry, \textit{A}, and $\chi^{2}$ 
-- that the the accuracy is largely independent of the vertically asymmetric matter distribution. This is likely due to the relatively shallow vertical metallicity gradient within each volume \citep[with an average slope across all volumes of $\bar{m}$ = -0.15 dex kpc$^{-1}$ compared with the MW's $m$=-0.3 dex kpc$^{-1}$,][]{Ivezic2008, Hayden2014, Imig2023}, meaning asymmetry in the density distribution does not manifest strongly in the mean abundance, which, because of the weak gradient, is fairly uniform across the entire volume.

In the middle left panel, we now consider the strength of the vertical metallicity profile (shown in Figure~\ref{fig:FigureA} in Appendix \ref{sec:appendixa} for each volume individually). Here, m is the slope of the best-fit line to the vertical [Fe/H] profile (shown as translucent black lines in Figure~\ref{fig:Figure3} and quantified for each solar volume in Figure \ref{fig:feh_gradient}). We find a weak negative correlation (Pearson's r=-0.24) between the strength of the vertical metallicity gradient and $\chi^{2}$: as the magnitude of the slope decreases, the $\chi^{2}$ increases. We conclude that, while a stronger gradient is ideal, OTI can still perform well as long as there exists a gradient with a slope steeper than -0.136 dex kpc$^{-1}$ (Figure~\ref{fig:feh_gradient}).

In the middle right panel, we plot the scale height as a function of $\Sigma_{\odot}$, the total surface mass density at $z$=1.1 kpc. We find a moderately strong negative correlation (Pearson's r=-0.49) between $\Sigma_{\odot}$ and $\chi^{2}$, indicating the importance of being well-sampled out to at least $z$=1.1 kpc. Since OTI leverages aggregate data to make statistical inferences, it is necessary to have enough stars (but $\Sigma_{\odot}$ also includes gas and dark matter), in a concentrated configuration about their midplane, in order to accurately predict \(\langle\)[Fe/H]\(\rangle\) and, consequently, potential parameters, like $a_{z}$. 

Finally, in the rightmost panel, we compute the median age of all stars in each volume. Figure~\ref{fig:ages} in Appendix \ref{sec:appendixe} displays the stellar age distribution within each volume. We find that \textit{OTI works best for stellar populations on disk-like orbits}, as both a higher median age and large scale height correlate with high $\chi^{2}$, or a decrease in the OTI model's accuracy (see Appendix~\ref{sec:appendixg} for a discussion of these properties related to OTI performance for each solar volume). This trend is because older stars follow more eccentric orbits than younger stars \citep{Spitzer51, Lacey84, Bird21}, possibly contributing to the large scale heights we measure. One of the primary assumptions in the OTI framework is separability of $R-z$ coordinates, but for kinematically-heated orbits, like those belonging to many older stellar populations, the epicyclic approximation breaks down \citep{BT2008}. 

\subsection{SIDM vs.~CDM Using $da_{z}/dz$}
\label{sec:da_dz_profiles}
Here, we discuss if results inferred from OTI hold discriminating power between two proposed models of dark matter. 
\citet{Arora2024} demonstrated that the gradient of the $a_z$ profile has distinguishing power between identical simulated galaxies run in a SIDM and CDM framework. 
Specifically, in the idealized limit, a substantially steeper (by 10$\%$ - 30$\%$) vertical acceleration gradient is predicted by SIDM than that predicted by CDM \citep[Figure 3,][]{Arora2024}. Given OTI's ability to measure the $a_z$ profile across a broad range of heights, we consider if, given OTI's errors, we can provide meaningful constraints to assess these competing dark matter models against.

While the uncertainties we report in our $\frac{da_{z}}{dz}$ profiles (Figure~\ref{fig:dadz}) are more constrained than those from other methods like Jeans modeling or binary pulsar acceleration measurements \citep[see Figure 3 from][for an updated constraint of 10$^{3.56 \pm 0.11}$ Gyr$^{-2}$]{Chakrabarti21, Donlon24}, it is important to note that the volume-to-volume variation is of the same order as the dark matter model dependence. This means that distinguishing between SIDM and CDM halos could be difficult, since we may not know whether the observed variations in the $a_z$ gradient profiles are due to differences across our sampled volumes or to the distinct imprint of SIDM on the gradient. We note the results presented in this paper are derived using a $\Lambda$CDM cosmology, and within this framework, we find a range of gradient profile slopes for both the FIRE simulated ground truth and the OTI predictions. This range overlaps with values reported for SIDM gradient profiles, signaling a need to exercise caution when interpreting these results and emphasizing the importance of further studies that incorporate multiple dark matter models.

\subsection{Comparison with Initial OTI Results}

In this section, we discuss our results in context with earlier OTI analysis, including the original study by \cite{APW21}, which introduced the OTI framework.  As discussed in \S\ref{sec:intro}, \cite{APW21} showed that element abundance distributions could be used to map out the vertical orbit structure without relying on second moments of the velocity distribution or knowledge of the selection function. They assumed a four-component analytic model for the mass model of the MW, computing actions and angles for this model using the Stackel-fudge implementation in \texttt{galpy} \citep{bovy15}. They minimized the angle-independence of the mean abundance deviation to infer the best-fit mass model for APOGEE-Gaia data, resulting in disk mass, scale height, and disk-halo mass ratio estimates with few-percent level precision. In turn, our results validate OTI’s efficacy to infer potential parameters and explore its sensitivity to properties of the local disk. We find, from our application of OTI in a realistic cosmological context, that the OTI-derived vertical acceleration profiles match the FIRE-2 profiles within 1$\sigma$ for 75$\%$ of solar volumes, demonstrating its promise for accurate potential parameter inferences, such as the MW estimates presented in \citet{APW21}. Here, we further show OTI performs best in regions characterized by low scale heights and young stellar populations, supporting its suitability for environments resembling the MW’s disk, such as the region studied in \citet{APW21}.

The next generation of OTI analysis refined the mass modeling to be more adaptable and further honed predictions for the MW. \citet{Horta2024} fit flexible representations of orbital shapes to the mean abundance distribution, a procedure we similarly follow, finding a total surface mass density of $\Sigma_{\odot}$($z$=1.1 kpc)=72$^{+6}_{-9}$ M$_{\odot}$ pc$^{-2}$. This result is compelling as it falls directly within the range of the OTI $\Sigma_{\odot}$ estimates we derived for similar regions, in which we demonstrate alignment with the true FIRE-2 potential parameters ($a_{z}$ and $\Sigma_{\odot}$) within 3$\sigma$ for 94$\%$ of volumes. This agreement within the simulations reinforces the accuracy of OTI in recovering true potential parameters, demonstrating its reliability as a dynamical mass modeling technique for MW estimates, such as those presented in \citet{Horta2024}. 

Subsequently, \citet{pricewhelan2024datadriven} applied this method to more complex systems, including the M1 pure $N$-body simulation from the \texttt{SMUDGE} suite \citep{Hunt2021}. This system was a controlled testing ground for OTI's adaptability given its non-axisymmetric kinematic structures \citep[see Figure 3 from][]{Hunt2021}. Importantly, M1 simulates the merger of an 8 $\times$ 10$^{10}$ M$_{\odot}$ dwarf galaxy into a 6 $\times$ 10$^{11}$ M$_{\odot}$ disk galaxy with a live 8.8 $\times$ 10$^{8}$ particle NFW DM halo with a mass resolution of 640 M$_{\odot}$, a 2.2 $\times$ 10$^{7}$ particle Hernquist bulge with a mass resolution of 540 M$_{\odot}$, and a  2.2 $\times$ 10$^{8}$ particle exponential disk and with a mass resolution of 160 M$_{\odot}$. For complete details of the initial condition and simulation setup, see \citep{Hunt2021}. 

M1 is evolved for 8.292 Gyr with a selected present-day snapshot of t = 6.874 Gyr to match the simulated satellite's position with observational values from \citep{VB20}. The advantage of M1 is the live disk with high particle resolution; like the analysis presented in this paper, \citep{pricewhelan2024datadriven} considers OTI's effectiveness at only one moment in time. Specifically, they applied OTI to one small (1 kpc$^{2}$) region of the disk, finding OTI accurately recovered $a_{z}$ near the midplane ($|z|<$ 0.5 kpc), with deviations from the true values of 15$\%$ at larger heights ($|z|$$\sim$1 kpc). While they did not vary the proximity to axisymmetric structure, they found that OTI still provided reasonable estimates  for a single volume.

In this work, we apply OTI to a range of solar volumes in a FIRE-2 simulated galaxy, m12i, which includes various feedback mechanisms (discussed in further detail in \S\ref{sec:simulations}) such as stellar feedback from radiation pressure, stellar winds, photoionization, photoelectric heating, and supernovae. The simulation we analyze was run in a fully cosmological context, but is in a recently quiescent state \citep[with the last major merger occurring at 9.86 Gyr, Table 4 from][]{Ansar25}. The offsets reported in \cite{pricewhelan2024datadriven} are comparable to the percent differences in the $a_{z}$ profiles that we calculate. As discussed in \S\ref{sec:a_z}, the average median percent difference below the midplane is 15$\%$ and 14$\%$ above the midplane, demonstrating the robustness of the OTI inferences in a more realistic, cosmological simulation that includes feedback over a range of solar volumes.

\section{Conclusions} 
\label{sec:conclusion}
Orbital Torus Imaging (OTI) \citep{APW21, pricewhelan2024datadriven, Horta2024} is a novel technique to infer parameters of the galactic gravitational potential using stellar ``labels", (here \(\langle\)[Fe/H]\(\rangle\)), dynamical invariants that trace out the orbit structure, as a function of vertical phase-space coordinates, $z$ and $v_{z}$. 
In this paper, we assess the reliability of this technique applied to m12i, a realistic cosmologically-derived MW-mass hydrodynamic galaxy simulation generated using the FIRE-2 code.
A synopsis of our work includes: 

\begin{itemize}
 
    \item We find agreement within 1\(\sigma\) between the OTI-inferred $a_{z}$ profiles and the known FIRE-2 profiles for 12 out of 16 solar volumes and agreement within 3\(\sigma\) for 15 out of 16 volumes (Figure~\ref{fig:Figure6}). Naively, if statistical uncertainties dominated, we would expect 11 out of 16 volumes to agree within 1\(\sigma\) and nearly all within 3\(\sigma\), which is almost what we observe. This suggests that systematic errors from the cosmological history and local asymmetry do not significantly dominate the error budget. In this study, we find the OTI a$_{z}$ inferences are $\sim85\%$ more precise than the Jeans modeling estimates.
    
    \item For $\Sigma_{\odot}$, we find agreement within 1\(\sigma\) for 11 out of 16 volumes and within 3\(\sigma\) for 15 out of 16 solar volumes (Figure~\ref{fig:Figure8}). Interestingly, the MW OTI-estimate \citep{Horta2024}, derived from APOGEE and Gaia data, lies within the range of densities we probe.
    
    \item We find OTI works well for a range of vertically-asymmetric distributions as well as vertical metallicity gradient strengths, demonstrating it is robust to these structural variations.
    
    \item OTI's accuracy declines for solar volumes with large scale heights and older stellar populations (Figure~\ref{fig:Figure9}). These two properties are correlated (Pearson's r=0.66, Figure~\ref{fig:pccm}) and indicate a shift away from kinematically cold, disk-like orbits. In contrast, OTI is most reliable for volumes that are more vertically concentrated (i.e., have low scale heights), are well-sampled out to $z$=1.1 kpc, and contain younger stars. This suggests OTI works best for dynamically cooler stellar populations, such as those that comprise the thin disk.
\end{itemize}

Our application of OTI in a benchmark FIRE-2 galaxy, m12i, lays the framework for future implementations of OTI on MW-like FIRE-2 galaxies with a range of merger histories. As a result of quantifying the accuracy of OTI performance in a recently quiescent, realistic simulated galaxy, we are equipped with a reference point to compare the deviation from truth of other, more drastically perturbed, FIRE-2 simulations. In this preliminary exploration of OTI's sensitivities to factors such as extended scale heights and relatively low density at $z$=1.1 kpc, we enable subsequent investigations of the method's biases in the presence of disequilibrium effects from merger events, enhancing its reliability in recovering Galactic potential parameters in our Universe.

\section*{acknowledgments} 
\label{sec:ack}
MO acknowledges funding support from the Faculty Mentor Program Fellowship of the University of California, Merced Graduate Division, and the University of California President’s Pre-Professoriate Fellowship. MO thanks the LSST Discovery Alliance Data Science Fellowship Program, which is funded by LSST Discovery Alliance, NSF Cybertraining Grant $\#$1829740, the Brinson Foundation, and the Moore Foundation; her participation in the program has benefited this work. SL acknowledges support from NSF grant AST-2109234 and HST grant AR-16624 from STScI.
This project was started at the Big Apple Dynamics
School (BADS) hosted by the Flatiron Institute July –
August 2021.

We generated simulations using: XSEDE, supported by NSF grant ACI-1548562; Blue Waters, supported by the NSF; Frontera allocations AST21010 and AST20016, supported by the NSF and TACC; Pleiades, via the NASA HEC program through the NAS Division at Ames Research Center.

FIRE-2 simulations are publicly available \citep{Wetzel2023} at \url{http://flathub.flatironinstitute.org/fire}.
Additional FIRE simulation data is available at \url{https://fire.northwestern.edu/data}.
A public version of the \textsc{Gizmo} code is available at \url{http://www.tapir.caltech.edu/~phopkins/Site/GIZMO.html}.

\vspace{2 mm}
\noindent{\emph{Software:} IPython \citep{ipython}, Matplotlib \citep{matplotlib}, Numpy \citep{numpy}, Scipy \citep{scipy}, \texttt{halo\_analysis} \citep{haloanalysis}, \texttt{gizmo\_analysis} \citep{gizmoanalysis}, \texttt{torusimaging} \citep{apw_torusimaging}, \texttt{JAX} \citep{bradbury2018}, \texttt{BlackJAX} \citep{cabezas2024blackjax}, \texttt{astropy} \citep{Astropy18}.

\bibliography{main}{}
\bibliographystyle{aasjournal}
\appendix
\counterwithin{figure}{section}
\FloatBarrier
\section{Vertical Metallicity Gradients in m12i}
\label{sec:appendixa}

In this section, we discuss the importance of the vertical metallicity gradient in the OTI modeling framework. Figure~\ref{fig:FigureA} presents a detailed view of the vertical metallicity gradient trends, similar to Figure~\ref{fig:Figure4}, but for each solar volume. We confirm the robustness of the smooth, systematic transition in \(\langle[\text{Fe/H}]\rangle\) from low $|z|$ and $|v_{z}|$ to high $|z|$ and $|v_{z}|$ in all volumes. See \S\ref{sec:application} for further details, to which this expanded figure provides additional context. The strength of this signal is quantified in Figure~\ref{fig:feh_gradient}. For all volumes, we report a measurable vertical metallicity gradient slope, ranging from $m$=-0.136 dex kpc$^{-1}$ to  $m$=-0.168 dex kpc$^{-1}$. The existence of a vertical metallicity gradient is a crucial component in OTI modeling, equipping the model with constraining power by being able to map more distinct orbital paths to the vertical distribution. While OTI necessitates a measurable gradient, its strength correlates only weakly with goodness-of-fit. For further details, see \S\ref{sec:disc} and Figure~\ref{fig:pccm} in Appendix \ref{sec:appendixf}.

\begin{figure*}[htbp]
\centering
\includegraphics[width=1\textwidth]{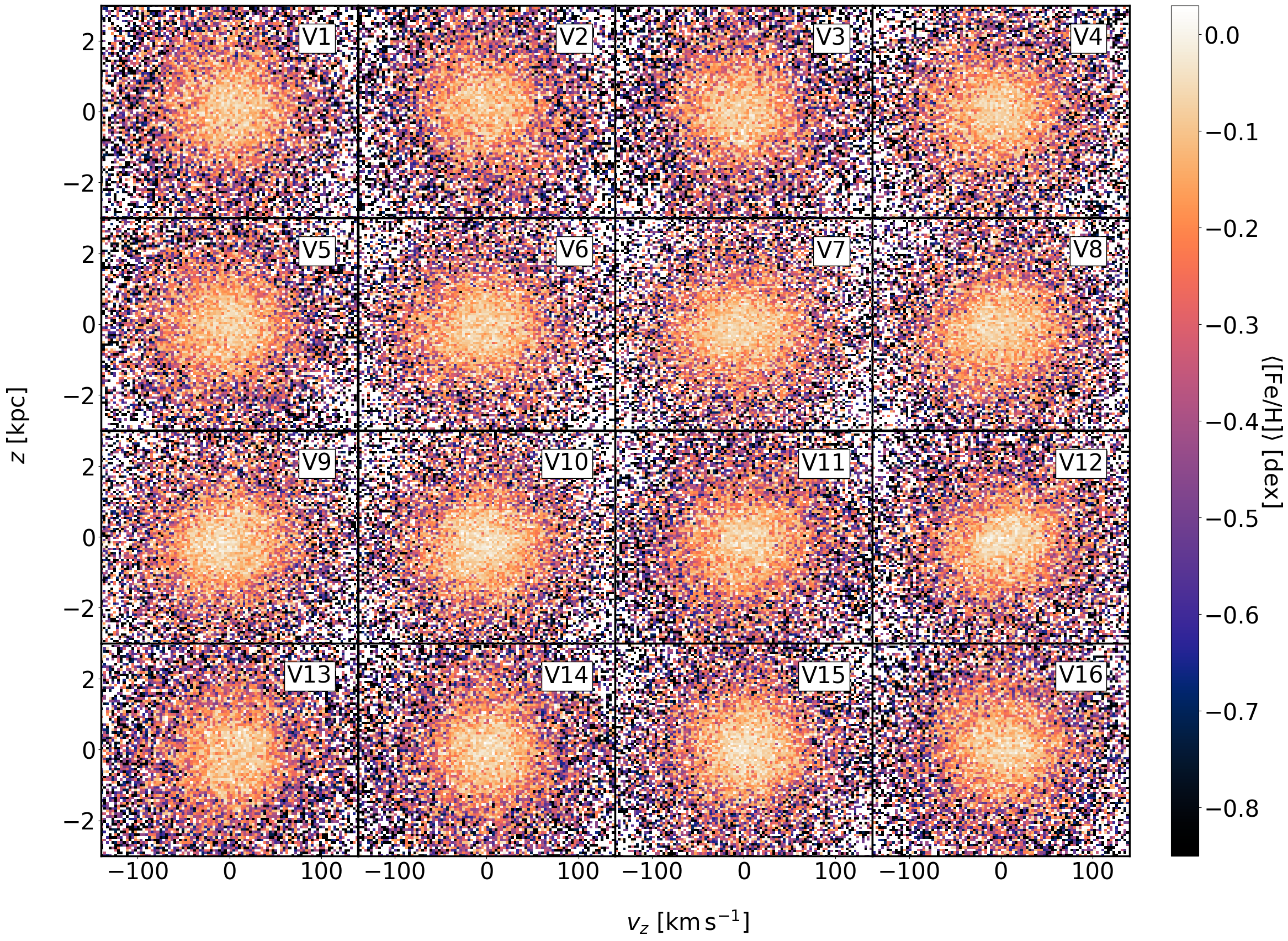}
\caption{The systematic transition in \(\langle[\text{Fe/H}]\rangle\) values across $z$ and $v_z$ within solar volumes provides strong constraints for OTI modeling. The 2D histogram shows ⟨[Fe/H]⟩ in vertical phase-space for stars in the 16 solar volumes described in Figure~\ref{fig:Figure1}. As in Figure~\ref{fig:Figure4}, the range for bin sizes is 170-226 (pc $\times$ km s$^{-1}$) due to the scatter within volumes. Each volume reveals a smooth gradient from higher \(\langle[\text{Fe/H}]\rangle\) values (orange) at lower $|z|$ and $|v_{z}|$ to lower \(\langle[\text{Fe/H}]\rangle\) values (purple) at higher $|z|$ and $|v_{z}|$, generating the basis for the OTI model's constraining power.
\label{fig:FigureA}}
\end{figure*}

\begin{figure*}[htbp]
\centering
\includegraphics[width=1\textwidth]{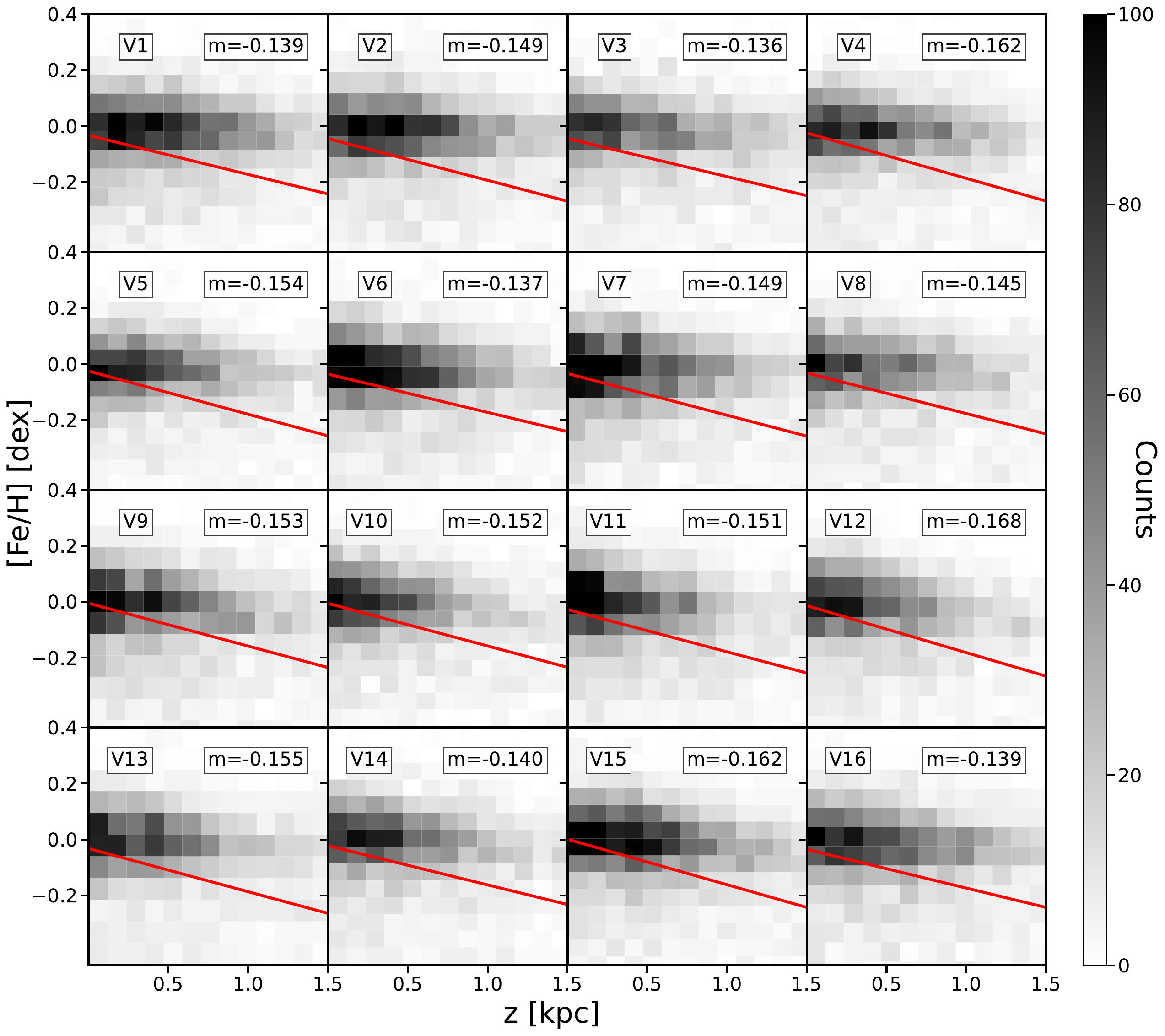}
\caption{As in the rightmost panel of Figure 1 from \citet{pricewhelan2024datadriven}, but for [Fe/H] instead of [Mg/Fe], we plot the mean abundance as a function of a proxy for $z_{\rm{max}}$. Our sample is drawn from the original stellar sample, contained within the solar volumes described in Figure~\ref{fig:Figure1}, with an additional cut on $v_{z}$ such that $|v_{z}|$ $<$ 10 km s$^{-1}$. Selecting for particles with low velocities enables us to keep only the stars that are near their apocenters, where they reach their highest vertical excursions. Across all solar-analog volumes, we note a nonzero, measurable gradient for the vertical metallicity profiles and include the slope of the best-fit line. OTI necessitates a measurable vertical metallicity gradient, as it provides the modeling with constraining power by being able to map distinct orbital paths to the vertical distribution. We compute a Pearson correlation coefficient between $|$m$|$ and $\chi^{2}$ and report a value of -0.24, indicating a weak correlation. The negative sign indicates the two variables move in opposite directions: as the slope decreases, $\chi^{2}$ climbs, indicating a worse fit corresponds to a shallower gradient. However, in this study, all 16 solar volumes exhibit a gradient sufficient for OTI analysis. 
\label{fig:feh_gradient}}
\end{figure*}

\clearpage
\FloatBarrier
\section{Vertical Density}
\label{sec:appendixb}
This section offers complementary insights into the vertical distribution and asymmetry of the stellar, gas, and dark matter, and total distribution across our sample of solar volumes (Figure~\ref{fig:vertical_density_profiles}) as well as a global perspective on the stellar distribution (Figure~\ref{fig:smd_2dhist}). Figure~\ref{fig:vertical_density_profiles} displays the vertical density profiles for stars, gas, dark matter, and the total distribution at a fixed galactocentric radius of 8 kpc. These profiles are computed by summing the mass within each solar volume and normalizing by the cylindrical volume element. They reveal a diversity of vertical asymmetries across the disk. However, as discussed in Figure~\ref{fig:Figure9} in \S\ref{sec:disc}, these vertical asymmetries do not significantly correlate with the accuracy of the $a_{z}$ estimates. This is because the package for OTI modeling, \texttt{torusimaging} \citep{pricewhelan2024datadriven}, recenters the midplane according to the peak density within each volume, mitigating the impact of vertical distortions on the accuracy of the model. 

\begin{figure*}[htbp]
\centering
\includegraphics[width=1\textwidth]{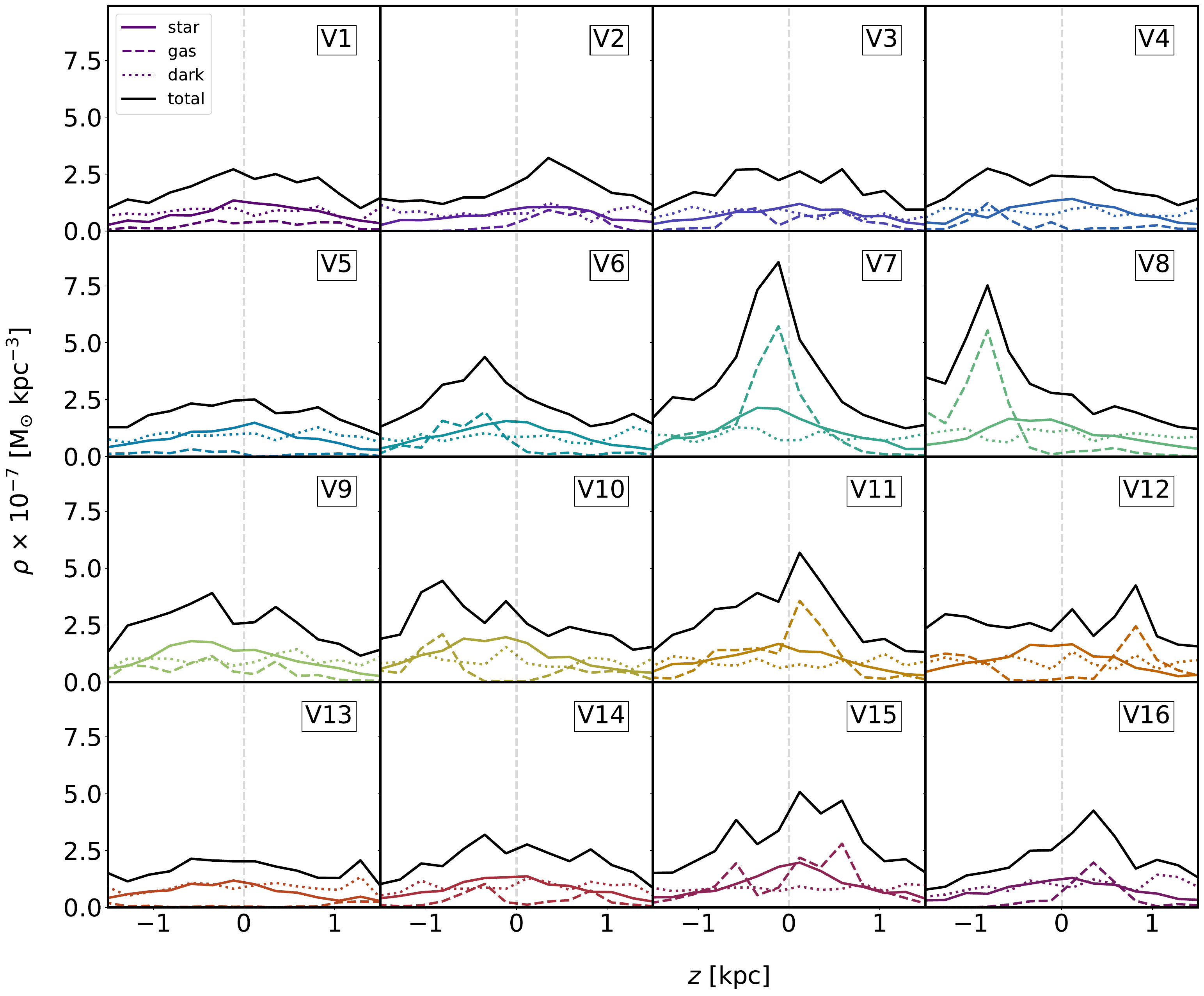}
\caption{Vertical density profiles show distinct distributions for stars, gas, dark matter, across all volumes. Profiles are shown at R = 8 kpc  for stars (colored solid lines), gas (colored dashed lines), dark matter (colored dotted lines), and all particles (black solid line). We compute the mass density by summing the mass within each solar volume (using bins of size 22 pc$^{2}$) and normalizing by the cylindrical volume element. Although vertical asymmetry is present in certain volumes, it does not correlate with the accuracy of $a_{z}$ estimates (see Figure~\ref{fig:Figure9} and \S\ref{sec:disc} for further discussion) because \texttt{torusimaging} \citep{pricewhelan2024datadriven} recenters the midplane according to the peak density within each volume.
\label{fig:vertical_density_profiles}}
\end{figure*}

Figure~\ref{fig:smd_2dhist} displays a 2D histogram of the stellar surface mass density in the $R-z$ plane for each volume, showing the stellar spatial distribution. The shaded region marks the radial extent of each solar volume. In spite of the asymmetries apparent in the edge-on stellar distributions, the $a_{z}$ accuracy is largely unaffected. These findings reinforce the results shown in Figure~\ref{fig:Figure9}, particularly that the presence of asymmetric conditions does not compromise OTI's accuracy, due to the automatic recentering done by \texttt{torusimaging}. Even though there is clear vertical asymmetry in both the stellar and total density distribution, it does not impede OTI’s efficacy in recovering potential parameters.

\begin{figure*}[htbp]
\centering
\includegraphics[width=1\textwidth]{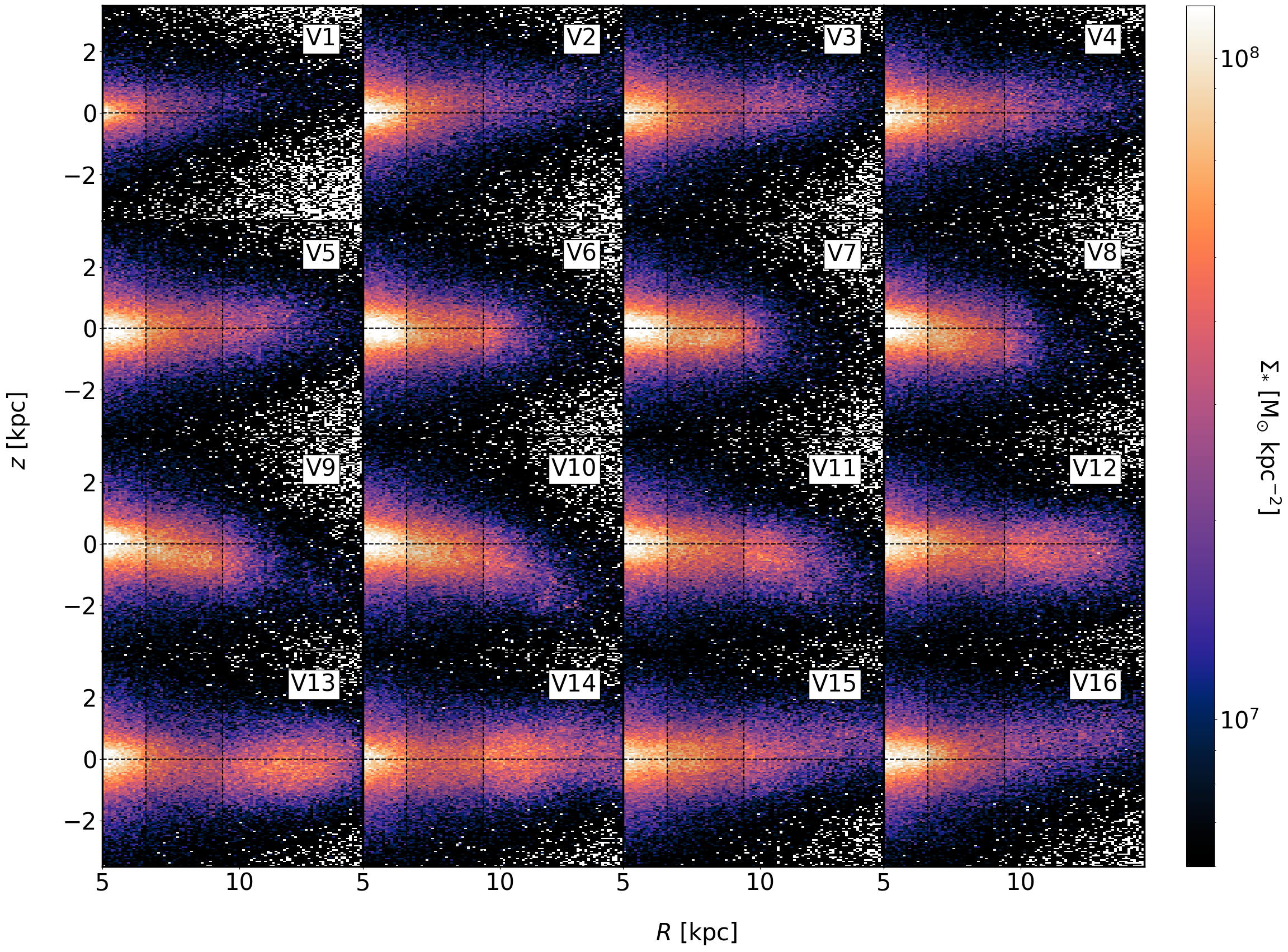}
\caption{2D histogram of stellar surface mass density in $R-z$ for each volume. The shaded panel indicates the radial extent of each solar volume. In spite of the diversity of asymmetrical conditions present in the edge-on stellar surface mass density distributions, the OTI model $a_{z}$ accuracy remains largely independent of this quantity. We note varying asymmetry about the midplane as a function of galactocentric radius for all volumes and plot the stellar mass normalized by bin area using bins of size 5 pc$^{2}$. See \S\ref{sec:disc} for further discussion.
\label{fig:smd_2dhist}}
\end{figure*}

\FloatBarrier
\section{Vertical Acceleration Gradient Profiles}
\label{sec:appendixd}
This section presents vertical acceleration gradient profiles across m12i's disk (Figure~\ref{fig:dadz}), comparing with work done by \citet{Arora2024}, which is discussed in \S\ref{sec:da_dz_profiles}. While OTI recovers this profile within the 3$\sigma$ limit for the majority of volumes, the intra-disk variation is of order the expected variation between SIDM and CDM. 

\begin{figure*}[htbp]
\centering
\includegraphics[width=0.92\textwidth]{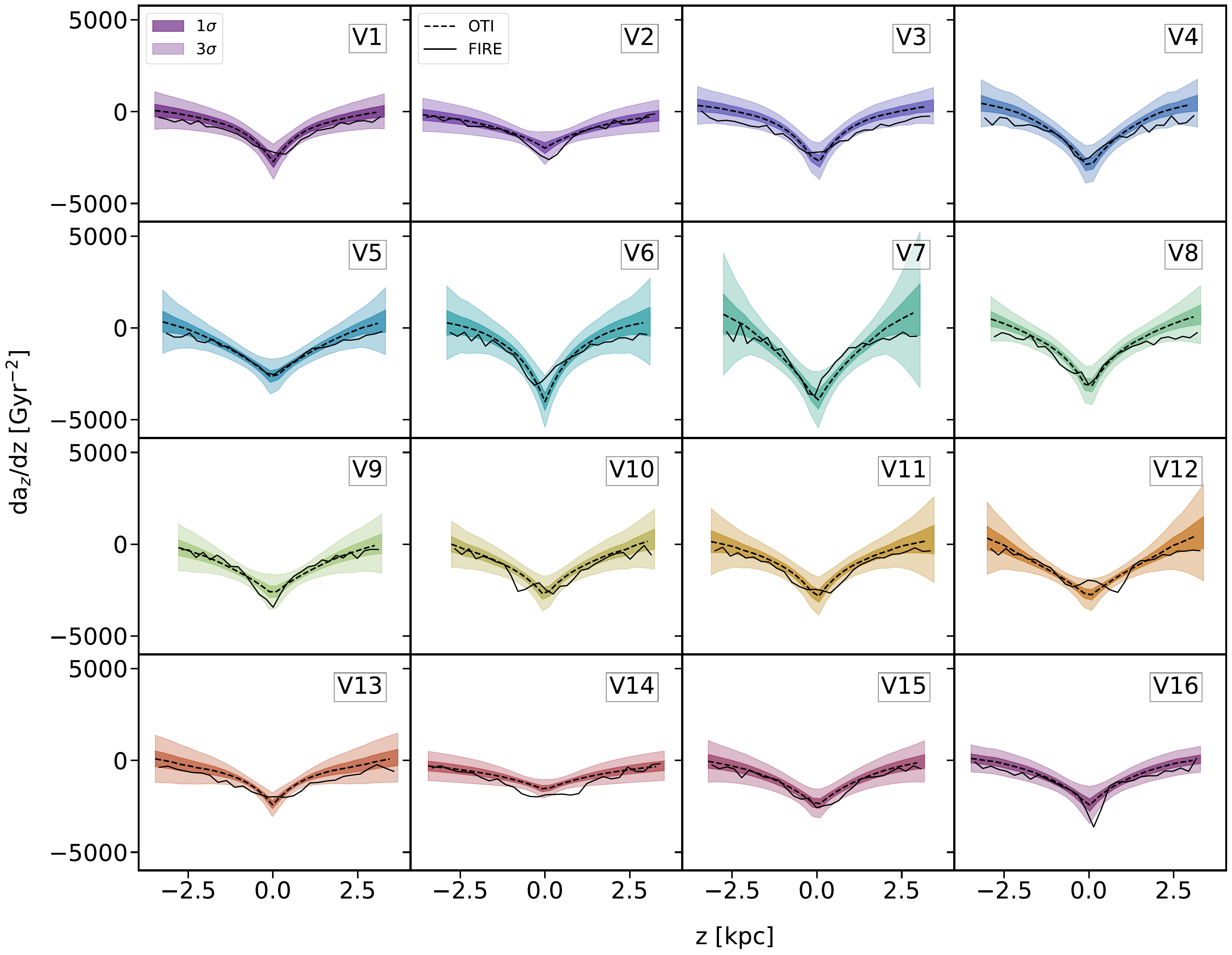}
\caption{Vertical acceleration gradient ($\frac{da_{z}}{dz}$) for all 16 solar analog volumes, similar to Figure 3 from \citet{Arora2024}, for m12i assuming a $\Lambda$CDM cosmology. For the majority of volumes, OTI (black dashed line) recovers the true FIRE (solid black line) $\frac{da_{z}}{dz}$ profile within the 3$\sigma$ limit. For reference, the MW's best-fit  $\frac{da_{z}}{dz}$ at $z$=0 kpc is -3200 $\pm$ 2560 Gyr$^{-2}$ \citep{Donlon24}, which lies within the range of values we probe with OTI. As discussed in \S\ref{sec:da_dz_profiles}, the volume-to-volume variation is on the same scale as the variation expected between SIDM and CDM, suggesting caution regarding its use in differentiating between dark matter models. See \S\ref{sec:da_dz_profiles} for further discussion.
\label{fig:dadz}}
\end{figure*}

\FloatBarrier
\section{OTI Model Hyperparameters}
\label{sec:appendixc}
Table \ref{tab:initial_model_properties} presents the parameters we set before optimizing the OTI model, including the knot locations for the label function and the Fourier distortion coefficient functions. Here, we also specify smoothing parameters for the spline interpolation. See \S\ref{sec:optimization} for a discussion on the optimization of the OTI model's 26 parameters.
 
\setlength{\tabcolsep}{4pt}
\begin{table*}[]
\caption{OTI model hyperparameters.}
\begin{center}
\begin{tabular}{|c||ccc|}
\hline

\textbf{Hyperparameters} & label & $e_{2}$ & $e_{4}$ \\ \hline \hline

$n_{\mathrm{knots}}$ & 8.0 & 10.0 & 5.0  \\ \hline

L2 regularization & 1.0 & 1.0 & 1.0  \\ \hline

Smoothing & 0.5 & 0.1 & 0.1  \\ \hline

Spacing power & 0.75 & 0.75 & 0.75  \\ \hline        

\end{tabular}
\end{center}
\tablecomments{Knots are used for spline interpolation, L2 regularization parameters prevent against overfitting the model, smoothing parameters balance the model fit and spline smoothness, and spacing power defines the spacing between spline knots.}
\label{tab:initial_model_properties}
\end{table*}

\clearpage
\FloatBarrier
\section{Stellar Ages}
\label{sec:appendixe}
Here, we present the distribution of stellar ages within each solar volume. In Figure~\ref{fig:ages}, vertical lines demarcate pre-disk, early disk, and late disk epochs of star formation according to definitions from \cite{McCluskey24}. As seen in Figure~\ref{fig:pccm} from Appendix \ref{sec:appendixf} and discussed in further detail in \S\ref{sec:disc}, the relative fraction of old-to-intermediate stars ($>$ 4 Gyr) correlates with the goodness-of-fit, $\chi^{2}$. This can be attributed to old stars being on non-disk-like orbits, violating the assumption of $R-z$ separability, as well as contributing to a shallower metallicity gradient.

\begin{figure*}[htbp]
\centering
\includegraphics[width=1\textwidth]{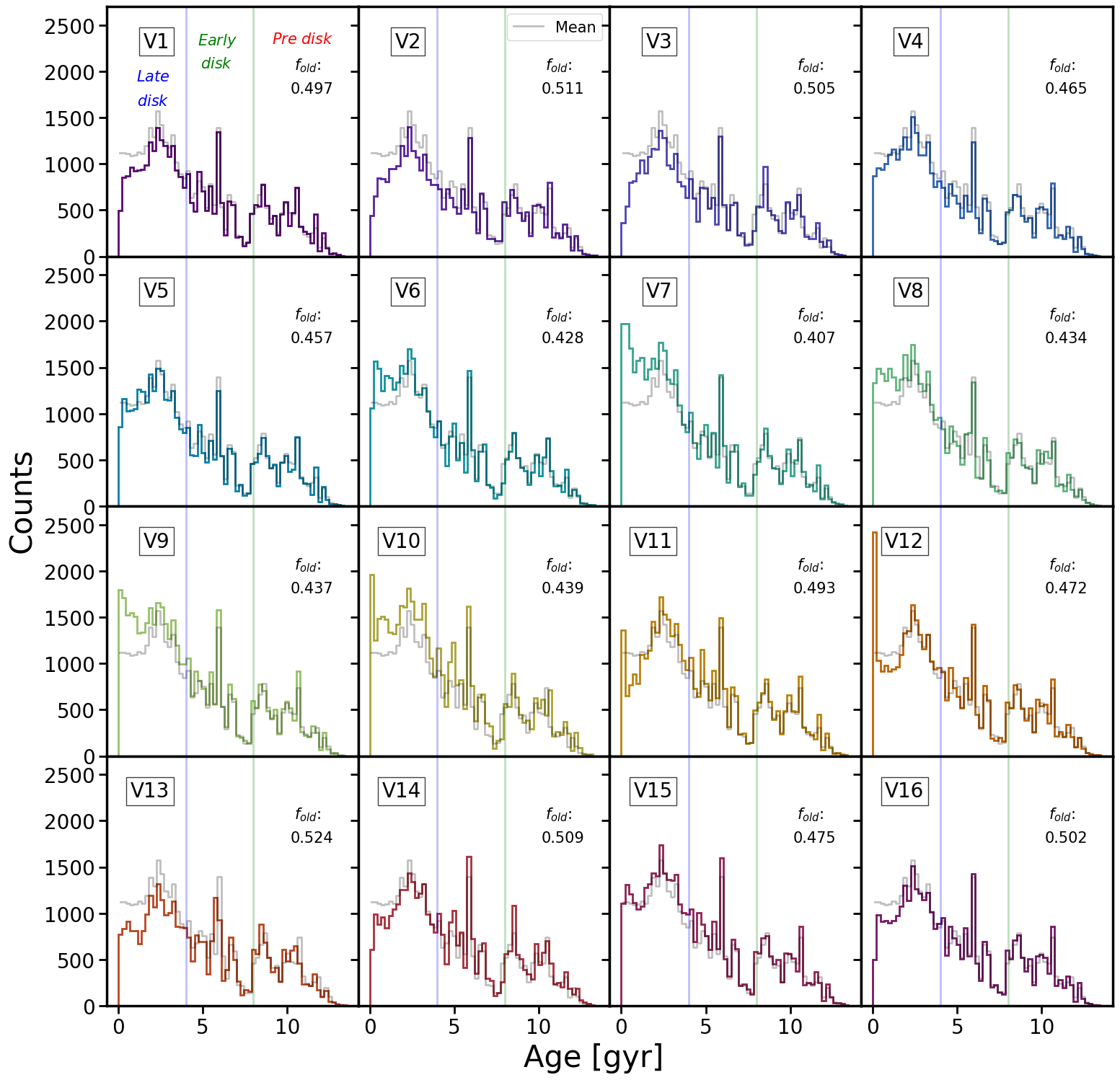}
\caption{Age histograms (colored lines) for each solar volume with mean histogram (gray line) across all volumes. Vertical lines indicate the pre-disk, early disk, and late disk definitions following \citet{McCluskey24}, denoting three different epochs of star formation. We find a moderately strong correlation (Pearson's r=0.56) between the fraction of intermediate-to-old ($>$ 4 Gyr) stars, f$_{old}$, and $\chi^{2}$ (Equation \ref{eq:chi}): volumes with a high ($>$ 0.5) fraction of old stars correlate with a loss in OTI model accuracy (i.e., high $\chi^{2}$) for the $a_{z}$ estimates. See \S\ref{sec:disc} for further discussion.
\label{fig:ages}}
\end{figure*}
\clearpage
\FloatBarrier
\section{Pearson Correlation Coefficients}
\label{sec:appendixf}
In this section, we present a quantification of the correlation (Figure~\ref{fig:pccm}) between the properties discussed in \S\ref{sec:disc}. We find that both vertical asymmetry in the density distribution and strength of the vertical metallicity gradient does not correlate strongly with goodness-of-fit, $\chi^{2}$, but total density about the midplane, median stellar age, and scale height all have moderately strong correlations with $\chi^{2}$, making these properties better indicators of the model's performance. See \S\ref{sec:disc} for further details.

\begin{figure*}[htbp]
\centering
\includegraphics[width=1\textwidth]{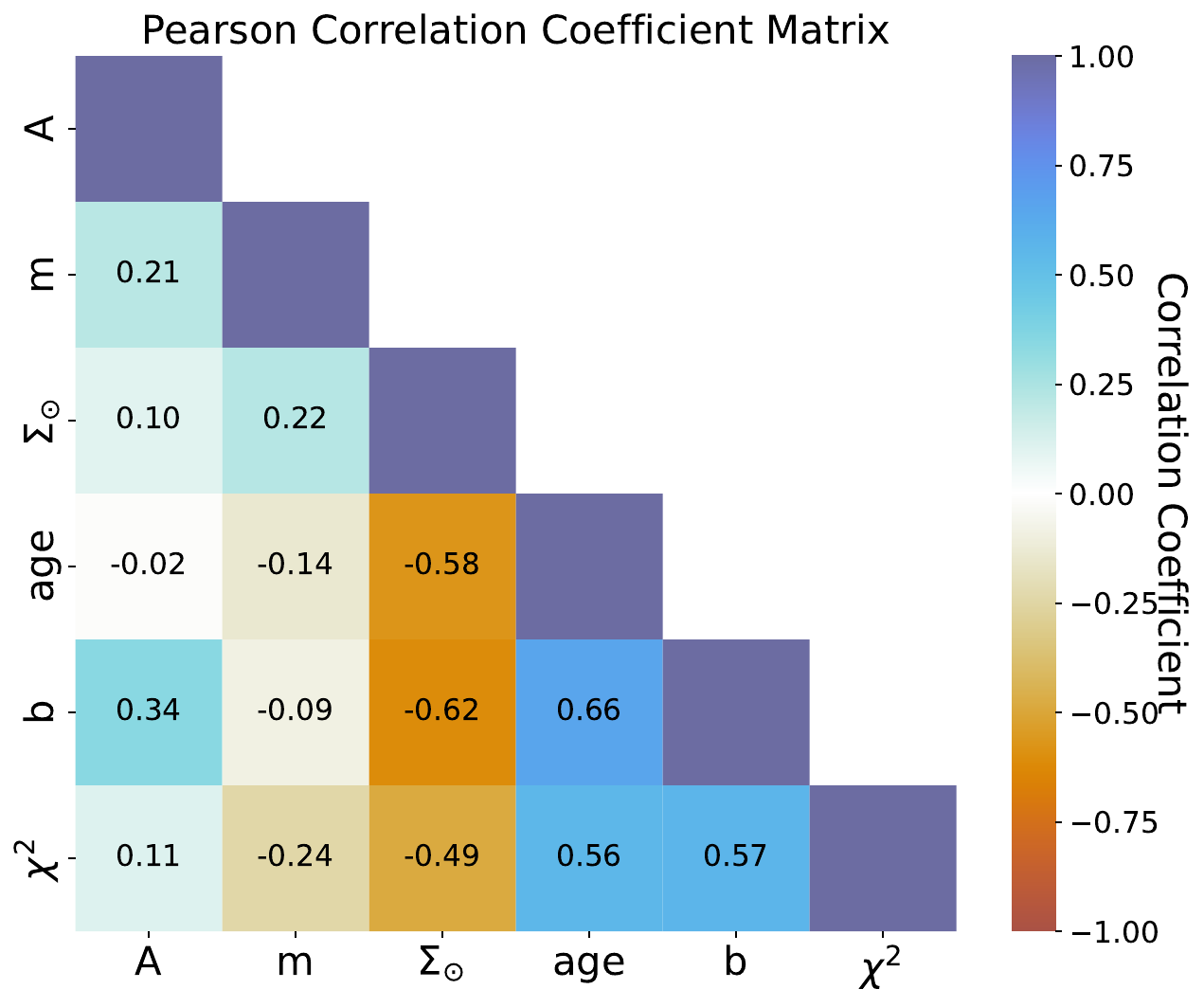}
\caption{Pearson correlation coefficient values for several aggregate statistics (per volume) considered in this work (from left to right or top to bottom): A, vertical asymmetry of the total (star, gas, and DM) distribution, m, slope of the \(\langle[\text{Fe/H}]\rangle\) vertical gradient, $\Sigma_{\odot}$, total density at a height of $z$=1.1 kpc, median stellar age, b, scale height, and $\chi^{2}$, goodness-of-fit (Equation \ref{eq:chi}). \S\ref{sec:disc} describes how these quantities are obtained. For illustrative purposes, we overlay a black rectangle highlighting the properties that have the highest correlations with $\chi^{2}$. The properties most correlated with our goodness-of-fit parameter, $\chi^{2}$, are: median stellar age, scale height, and total density, in the order of most to least correlated. Conversely, the properties which correlate very weakly with $\chi^{2}$ are vertical asymmetry and strength of the vertical metallicity gradient. 
\label{fig:pccm}}
\end{figure*}

\clearpage
\FloatBarrier
\section{Local Disk Properties}
\label{sec:appendixg}
Here, we summarize our findings of the correlations between several properties of the local disk and OTI model performance. Figure~\ref{fig:volume_metrics} presents values for vertical asymmetry, the slope of the vertical metallicity gradient, total density ($\Sigma_{\odot}$) at z=1.1 kpc, median age, scale height ($b$), and the goodness-of-fit parameter, $\chi^{2}$, normalized by the total number of parameters, $\theta$, for each solar volume. Each cell is color-coded based on relative score: the five worst scoring volumes are marked in red, the five best in green, and the remaining intermediate scoring volumes in yellow. Best and worst are defined in terms of scores that correlate with a loss in OTI accuracy. As discussed in Section \S\ref{sec:disc}, asymmetry and the slope of the vertical metallicity gradient only weakly correlate with normalized $\chi^{2}$, whereas total density, median age, and scale height have moderately high correlations. This suggests that the latter properties are more relevant for OTI accuracy. Figure~\ref{fig:volume_metrics} reinforces the trend seen in Figure~\ref{fig:Figure9}, that volumes with low $\Sigma_{\odot}$, large scale heights, and high median ages (e.g., V14, V13, and V3) tend to have the highest normalized $\chi^{2}$ values, indicating lower accuracy in $a_{z}$. In contrast, volumes with moderate to high total densities, younger stellar populations, and relatively small to moderate scale heights (e.g., V6, V7, and V9) exhibit the lowest normalized $\chi^{2}$, reflecting higher accuracy in the OTI $a_{z}$ inferences.

\begin{figure*}[htbp]
\centering
\includegraphics[width=1\textwidth]{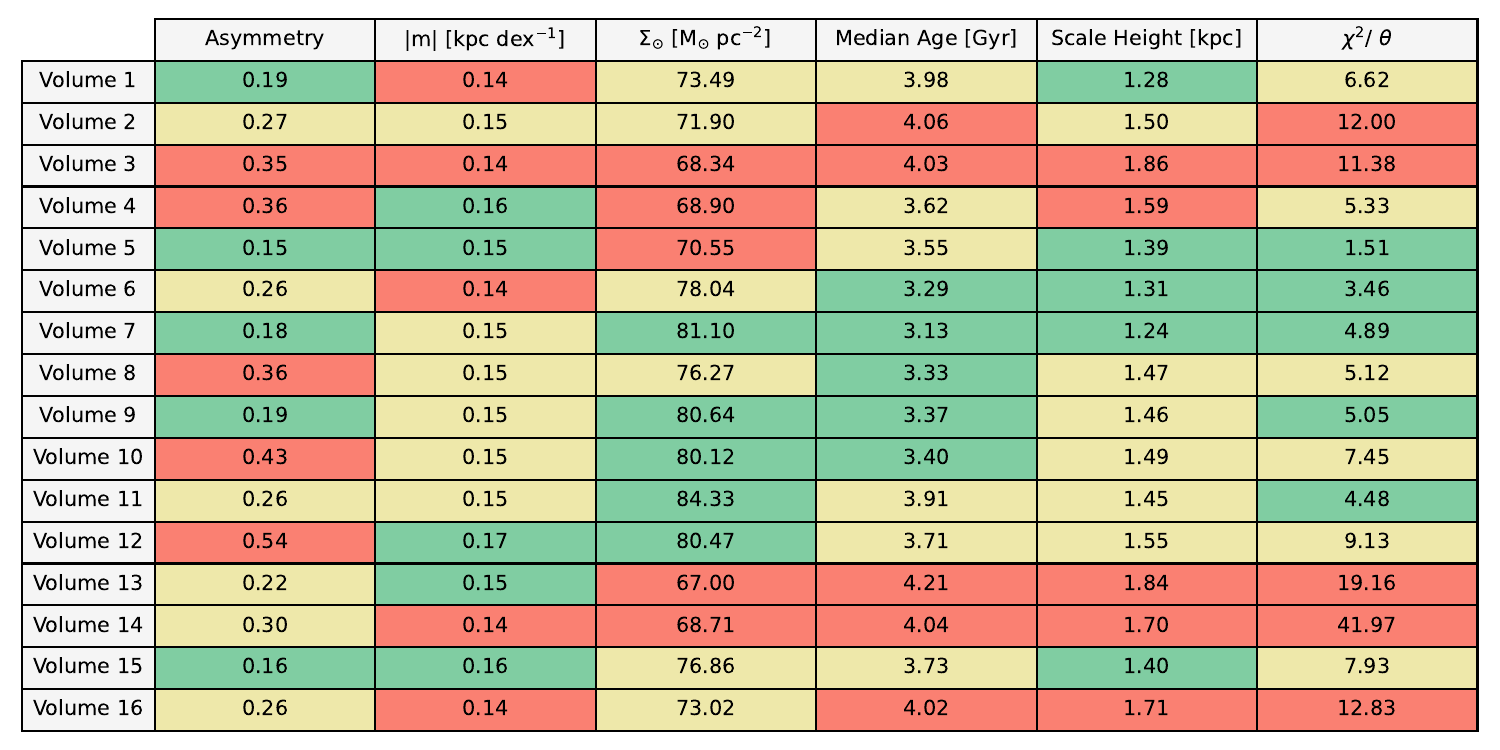}
\caption{This table summarizes key local disk properties across solar volumes, including asymmetry, vertical metallicity gradient slope, total density at z=1.1 kpc, median age, scale height, and normalized $\chi^{2}$. We color code each cell by relative score: red (worst), yellow (intermediate), and green (best); worse scores are those we find correlate with a decrease in OTI accuracy for the $a_{z}$ estimates. Total density, age, and scale height show the strongest correlation with normalized $\chi^{2}$, suggesting greater importance for OTI accuracy than asymmetry or vertical metallicity gradient slope (see \S\ref{sec:disc} and Figure~\ref{fig:Figure9} for further discussion). Volumes with high $\chi^{2}$ (e.g., V14, V13, V3) tend to have low $\Sigma_{\odot}$, large scale heights, and older populations, while those with low $\chi^{2}$ (e.g., V6, V7, V9) show the opposite trend.
\label{fig:volume_metrics}}
\end{figure*}

\end{document}